\documentclass[
    aps,
    prd,
    reprint,
    amsmath,amssymb,
    groupedaddress,
    twocolumn,showpacs,
    ]{revtex4-2}
\usepackage{graphicx}
\usepackage{dcolumn}
\usepackage{bm}
\usepackage[mathlines]{lineno}
\usepackage{multirow}
\usepackage{comment}
\usepackage{enumerate}
\usepackage{color}

\begin{document}


\title{Parameterization of Stochasticity in Galaxy Clustering and Reconstruction of Tomographic Matter Clustering}


\author{Shuren Zhou $^{1,2}$}
 \email{zhoushuren@sjtu.edu.cn}
\author{Pengjie Zhang $^{1,3,2}$}
 \email{zhangpj@sjtu.edu.cn}
\affiliation{
$^{1}$ Department of Astronomy, School of Physics and Astronomy, Shanghai Jiao Tong University, Shanghai, China \\
$^{2}$ Key Laboratory for Particle Astrophysics and Cosmology (MOE)/Shanghai Key Laboratory for Particle Physics and Cosmology, China\\
$^{3}$ Division of Astronomy and Astrophysics, Tsung-Dao Lee Institute, Shanghai Jiao Tong University, Shanghai, China
}


\date{\today}

\begin{abstract}
The stochasticity in galaxy clustering, the mismatch between galaxy and underlying matter distribution, suppresses the matter clustering amplitude reconstructed by the combination of galaxy auto-correlation and galaxy-galaxy lensing cross-correlation. In this work, we solve the stochasticity systematics by parameterizing the cross correlation coefficient $r(k)$ between galaxy and matter. We investigate the performance of 12 kinds of parameterization schemes, against the cosmoDC2 $\&$ TNG300-1 galaxy samples over a wide range of redshift and flux cut. The 2-parameter fits are found to describe the stochasticity up to $k_{\rm max}=0.9\,{\rm Mpc^{-1}}h$, while the best performing quadratic scheme $r^2_s(k) = 1+c_1 k+c_2 k^2$ reaches better than $1\%$ accuracy for both the direct ${r}^2_s(k)$ fit and reconstructing matter clustering. Then, we apply the accurate quadratic scheme to forecast the tomographic matter clustering reconstruction by the combination DESI-like LRG $\times$ CSST-like cosmic shear. Depending on assumption of stochasticity, we find that the neglect of a serious stochasticity would result in significant systematic bias in both the reconstruction and the inferred cosmological parameters, even if we adopt scale cut $k_{\rm max}=0.1\,{\rm Mpc^{-1}}h$. We demonstrate the necessity of including stochasticity in reconstruction, and forecast that the reconstruction alone enables a $S_8$ constraint at about $1.5\%$ precision, free from galaxy bias and stochasticity. We will validate our method for DESI spectroscopic survey, and the analysis is expected to be complementary to DESI cosmological constraint by BAO and RSD.

\end{abstract}


\maketitle


\section{Introduction \label{sec:intro}}

Galaxy clustering contains a wealth of cosmological information, and serves as an important probe to measure the large scale structure in the ongoing spectroscopic galaxy surveys such as DESI \cite{abareshi2022overview, adame2023validation, collaboration2023early} and LSST \cite{ivezic2019lsst}. The upcoming million spectroscopic objects offer precision but biased matter clustering measurement. Meanwhile, other cosmological observables, such as galaxy lensing delivered from Euclid satellite mission \cite{scaramella2022euclid} and the future CSST cosmic shear survey \cite{cao2018testing, gong2019cosmology, lin2022forecast}, is complementary to the galaxy clustering. The gravitational lensing is the projected distortion effect of the background object due to the deflection of arrival photons by the foreground matter clustering, providing the unbiased but integrated matter clustering amplitude. The joint analysis of both two, specifically, galaxy auto-power spectrum ${C}^{gg}_\ell$ and cross-power spectrum with lensing convergence ${C}^{g\kappa}_\ell$, can break the degeneracy between galaxy bias and matter clustering amplitude, obtaining the tomographic matter clustering power spectrum \cite{peacock2018wide, krolewski2021cosmological, white2022cosmological, marques2024cosmological}.

The galaxy deterministic bias can be cancelled model-independently combining different data set  (e.g. the gravitational potential decay rate measurement \cite{Zhang:2005gh,2022ApJ...938...72D} and $E_G$ statistic \cite{Zhang:2007nk}). This would motivate a simple estimator for the matter power spectrum, 
\begin{eqnarray}
    \label{equ:Pmm}
    \hat{P}_{mm}(k) = \hat{P}_{gm}^2(k) /\hat{P}_{gg}(k) \quad . 
\end{eqnarray}
Here $\hat{P}_{gg}$ and $\hat{P}_{gm}$ are the galaxy-galaxy and galaxy-matter 3D power spectrum, realized by the angular power spectrum $\hat{C}^{gg}$ and $\hat{C}^{g\kappa}$ with narrow redshift width. It seems a promising estimate of matter clustering, since any galaxy bias $b(k)$ spontaneously disappears if the dependence of galaxy on matter distribution is deterministic. However, apart from the deterministic galaxy bias, there are stochastic components in the galaxy-matter relation, arising from the intrinsic discrete property and the complex physical processes in halo collapse and galaxy formation \cite{desjacques2018large, bonoli2009halo, seljak2009suppress, baldauf2013halo}. Therefore, the ensemble average of the estimator is biased by a multiplicative $r^2$, 
\begin{eqnarray}
    \label{equ:Pmm_expect}
    \langle\hat{P}_{mm}(k)\rangle=r^2(k)P_{mm}(k) \quad .
\end{eqnarray}
Here $P_{mm}$ is the theoretical expectation, and $r^2$ is the cross correlation coefficient between galaxy number counts overdensity $\delta_g({\bf k})$ and underlying matter overdensity $\delta_m({\bf k})$. The function $r^2(k)$ quantifies the efficiency of galaxy tracing the matter distribution, and we always have $r(k)<1$ for stochastic tracers. The neglect of the impact of stochasticity in galaxy clustering will suppress the reconstructed matter power spectrum by $r^2(k)$, equivalently suppressing the amplitude by $r(k)$, depending on specific scale $k$ we interest in. The overall suppression also propagates to the cosmological parameters constraint and results in potentially systematics. In order to leverage the unbiased clustering information in non-linear scale, which is contaminated by stochasticity, we expect an accurate $r(k)$ parameterization to describe the suppression in estimator Eq.~(\ref{equ:Pmm}).


Two pathways have been explored to alleviate the stochasticity induced systematics. One pathway is to realize the reduced stochasticity by some proper weights. The weighting schemes are adopted by weighing $\delta_g$ with galaxy bias \citep{bonoli2009halo}, function of host halo mass \citep{seljak2009suppress, hamaus2010minimizing, cai2011optimal, liu2021biased} or some principal components \cite{Tegmark_1999, hamaus2012optimal, shurenzhou2023principal}. Especially, the principal component analysis (PCA) suggests that the principal eigenmode of galaxy clustering traces the matter clustering, while only two eigenmodes significantly deviate from the shotnoise expectation \cite{hamaus2012optimal, shurenzhou2023principal}. It implies that the effective degree of freedom (${\rm DoF}$) in stochasticity is finite and approximately ${\rm DoF}\simeq2$. Another pathway is to parameterize the stochasticity with some phenomenological parameters. In the perspective of effective field theory of large scale structure (EFTofLSS), the stochasticity contribution to the galaxy auto-/cross-power spectrum is expanded in even power of Fourier mode $k$ \cite{desjacques2018large, cabass2020likelihood, eggemeier2019bias, alexander2020testing, eggemeier2021testing, pezzotta2021testing, kokron2022priors, schmidt2016towards, perko2016biased, chudaykin2020nonlinear, ashley2016biased, moretti2023modified, mergulhao2023effective}.
In the widely adopted scheme, the expansion series in real space is truncated at the first two terms, where the first constant term characterizes the deviation of Poisson shot noise and the second $k^2$-scaled term describes the scale dependence. In the redshift space, an additional free parameter is required to describe the anisotropic directional dependence \cite{ashley2016biased, mergulhao2023effective}. These phenomenological descriptions are fully investigated in the content of galaxy bias expansion scheme, but it is difficult to ascertain the impact of stochasticity because of the mixture with higher order terms omitted in expansion truncation \cite{alexander2020testing, eggemeier2021testing, kokron2022priors}. Moreover, perturbation description only applies to the scale $k\lesssim 0.3\,{\rm Mpc^{-1}}h$ due to the strong non-linearity \cite{schmittfull2019modeling, alexander2020testing, eggemeier2021testing, pezzotta2021testing}, losing modeling ability beyond linear region. 

In this work, we aim to seek a parameterization for the cross correlation coefficient $r(k)$, following the perspective of EFTofLSS. In the following context, we investigate various parameterization forms and test their performance against the galaxy samples in simulations. The simple quadratic scheme $r_s^2=1+c_1 k +c_2 k^2$ is identified as the most promising parameterization. Then, we demonstrate the reconstruction of the matter power spectrum combining galaxy clustering and galaxy lensing, together with the best performing quadratic $r(k)$ scheme. Our method is suited for the tomographic clustering analysis in DESI Luminous Red Galaxies (LRG) samples.
DESI was designed to conduct a survey covering $14000\,{\rm deg^2}$ footprint, obtaining about $40$ millions of spectroscopic targets ranging from redshift 0 to $3.5$. These precision measurements will enable the constraint of $f\sigma_8$ to an aggregate precision $0.95\%$ by combining the baryonic acoustic oscillations (BAO) and redshift space distortion (RSD) \cite{adame2023validation}, where $f$ is the growth and $\sigma_8$ is the amplitude of matter fluctuation in spheres of $8\,h^{-1}{\rm Mpc}$. Our analysis enables the direct constraint on fluctuation amplitude $\sigma_8$, which could serve as a strong supplement for DESI's fiducial analysis on structure growth. It is also an immediate comparison with the results from cosmic microwave background (CMB) constraint \cite{aghanim2020planck}, to help to clarify whether the anomaly in low redshift observations compared to CMB is attributed to new physics or merely systematic contamination \cite{tutusaus2023first, krolewski2021cosmological, white2022cosmological}.

The paper is organized as followed. In Section~\ref{sec:parameterization_scheme}, we introduce the theoretical framework under the perturbation expansion perspective to illustrate the impact of stochasticity on the galaxy power spectrum, and then introduce our parameterization schemes. Readers with limited interest in the background can skip Section~\ref{sec:parameterization_scheme} and refer to Table~\ref{table:formula} directly. In Section~\ref{sec:methodology}, we introduce the methodology to test the performance of the parameterization and the methodology combining the galaxy clustering and galaxy lensing. In Section~\ref{sec:results}, we present results of the performance test. In Section~\ref{sec:desi-like}, we present the fisher forecast for the combination of DESI-like spectroscopic tracers and CSST-like cosmic shear.


\section{Parameterization Scheme \label{sec:parameterization_scheme}}

\begin{table*}
    \renewcommand{\arraystretch}{1.3}
\caption{\label{table:formula} Parameterization expressions of $r^2(k)$ and $r^2_s(k)$, defined in Eq.~(\ref{equ:r2}) and Eq.~(\ref{equ:r2_ss}). Notation $r^2_{(s)}$ refers to the parameterization that are applied to both $r^2$ and $r^2_s$. 
$P_{mm}$ and $P_{gg,s}$ are the matter power spectrum and the galaxy power spectrum subtracted the Poisson expectation. $\Delta^2(k) \equiv k^3 P(k) /(2\pi^2)$ is the corresponding variance.
}
\begin{ruledtabular}
\begin{tabular}{lclll}
    &Label & Formula & Free Parameters & \\ \hline
    & $\texttt{Expan-mm-1}$ & $r^2_{(s)} = \left[ 1+ \alpha\, /\left(\bar{n}_g P_{mm}\right) \right]^{-1} $ & $\;\alpha\;$  &  \\
    & $\texttt{Expan-mm-2}$ & $r^2_{(s)} = \left[ 1+ \alpha\,(1+\beta k^2 ) /\left(\bar{n}_g P_{mm}\right) \right]^{-1} $ & $\;\alpha,\; \beta \;$ & \\
    & $\texttt{Expan-mm-3}$ & $r^2 = \left[ 1+ \alpha\,(1+\beta k^2 +\gamma k^4) /\left(\bar{n}_g P_{mm}\right) \right]^{-1} $ & $\;\alpha, \;\beta, \;\gamma \;$ & \\
    & $\texttt{Expan-gg-1}$ & $r^2 = \left[ 1+ \alpha\, /\left(\bar{n}_g P_{gg,s}\right) \right]^{-1} $ & $\;\alpha\;$  & \\
    & $\texttt{Expan-gg-2}$ & $r^2 = \left[ 1+ \alpha\,(1+\beta k^2 ) /\left(\bar{n}_g P_{gg,s}\right) \right]^{-1} $ & $\;\alpha,\; \beta \;$  &  \\
    & $\texttt{Expan-gg-3}$ & $r^2 = \left[ 1+ \alpha\,(1+\beta k^2 +\gamma k^4) /\left(\bar{n}_g P_{gg,s}\right) \right]^{-1} $ & $\;\alpha, \;\beta, \;\gamma \;$   &  \\
    & $\texttt{Expan-new-2}$ & $r^2_{(s)} = (1+\beta k^2) \left[ 1+ \alpha/\left(\bar{n}_g P_{mm}\right) \right]^{-1} $ & $\;\alpha,\; \beta \;$  & \\
    & $\texttt{Zheng2013-mm}$ & $r^2_{(s)} = \left(1 +\beta \Delta^2_{mm}\right)^{-1} \left[ 1+ \alpha/\left(\bar{n}_g P_{mm}\right) \right]^{-1} $ & $\;\alpha,\; \beta \;$  & \\
    & $\texttt{Zheng2013-gg}$ & $r^2 = \left(1 +\beta \Delta^2_{gg,s}\right)^{-1} \left[ 1+ \alpha/\left(\bar{n}_g P_{gg,s}\right) \right]^{-1} $ & $\;\alpha,\; \beta \;$  & \\
    & $\texttt{Q-bias}$ & $r_s^2 = \left(1 +c_2 k^2\right)/\left(1+ c_1 k\right)$ & $\;c_1, \;c_2 \;$  & \\
    & $\texttt{Quadratic}$ & $r_s^2 = 1 +c_1 k + c_2 k^2$ & $\;c_1, \;c_2 \;$  & \\
    & $\texttt{Quadratic-1}$ & $r_s^2 = 1 + c_2 k^2$ & $\;c_2\;$  &
\end{tabular}
\end{ruledtabular}
\end{table*}

In Fourier space, the galaxy overdensity $\delta_g(\rm x)$ can be decomposed into the deterministic and the stochastic part
\begin{equation}
    \label{equ:delta_g}
    \delta_g({\bf k}) = b(k) \delta_m({\bf k}) + \delta_{\mathcal{S}}({\bf k})
\end{equation}
Here the scale dependent galaxy bias is defined as $b(k)\equiv P_{gm}(k)/P_{mm}(k)$, therefore $\langle\delta_m({\bf k})\delta_{\mathcal{S}}^*({\bf k})\rangle=0$. In the galaxy auto power spectrum, the stochasticity induces an additionally inconsistent term, that $P_{gg}(k) = b^2(k)P_{mm}(k) +P_{\mathcal{S}}(k)$. We can quantify the efficiency of the galaxy tracing the matter field by the cross correlation coefficient
\begin{eqnarray}
    \label{equ:r2}
    r^2(k) &\equiv\;& \frac{P_{gm}^2}{P_{mm}P_{gg}} = \frac{1}{1+ P_{\mathcal{S}} \left(b^2\,P_{mm}\right)^{-1} }  \quad .
\end{eqnarray}
Another widely adopted convention is 
\begin{eqnarray}
    \label{equ:r2_ss}
    r_s^2(k) \equiv \frac{P_{gm}^2}{P_{mm}\, P_{gg,s}}  \quad ,
\end{eqnarray}
where  the Poisson noise expectation is subtracted from the galaxy power spectrum, $P_{gg,s}(k) = P_{gg}(k)-1/\bar{n}_g $. In the following context, we utilize both $r^2$ and $r^2_s$ to describe stochasticity since they are equivalent in practical applications.

In the perspective of EFTofLSS, a complete description of the general bias expansion requires the stochastic variables due to the coupling of the small scale modes, statistic properties of which in general depend on the local observables \cite{mcdonald2009clustering, mirbabayi2015biased, desjacques2018large, schmittfull2019modeling}. Up to the third order, the analogous expansion of the stochastic components in Eq.~(\ref{equ:delta_g}) is
\begin{eqnarray}
\label{equ:stoch_expand}
    \delta_{\mathcal{S}}({\bf x}) &=\,& \varepsilon_{0}({\bf x}) + \varepsilon_{1}({\bf x}) \delta_m({\bf x}) + \varepsilon_{2}({\bf x}) \delta_m^2({\bf x}) + \varepsilon_{s^2}({\bf x}) s^2({\bf x}) \nonumber\\ 
    & +\,& b_{s^2} \left[ s^2_{\mathcal{S}}({\bf x}) - \langle s^2_{\mathcal{S}}({\bf x})\rangle \right] + b_{s^3} \left[ s^3_{\mathcal{S}}({\bf x}) - \langle s^3_{\mathcal{S}}({\bf x})\rangle \right]  \nonumber\\
    & +\,& b_{\delta s^2} \, \delta_ms^2_{\mathcal{S}}({\bf x}) 
    +\, b_{st} \left[ st_{\mathcal{S}}({\bf x}) - \langle st_{\mathcal{S}}({\bf x})\rangle \right]  \nonumber\\
    & +\,& b_{\psi} \left[ \psi_{\mathcal{S}}({\bf x}) - \langle \psi_{\mathcal{S}}({\bf x})\rangle \right] \;+\, \mathcal{O}(\varepsilon^4)  \quad ,
\end{eqnarray}
where the tidal field and other higher order terms are defined as \cite{mcdonald2009clustering}
\begin{eqnarray}
    s_{ij}({\bf x}) &=& \left(\nabla_i\nabla_j\nabla^{-2} - \frac{1}{3}\delta_{ij}^{K}\right)\delta({\bf x})  \nonumber\\
    t_{ij}({\bf x}) &=& \left(\nabla_i\nabla_j\nabla^{-2} - \frac{1}{3}\delta_{ij}^{K}\right)\left[ \theta({\bf x}) - \delta({\bf x}) \right]  \nonumber\\
    \psi({\bf x}) &=& \theta({\bf x}) - \delta({\bf x}) -\frac{2}{7}s^2({\bf x}) +\frac{4}{21}\delta^2({\bf x})   \quad .
\end{eqnarray}
Here the stochastic variables $\varepsilon_{0}, \varepsilon_{1}, \varepsilon_{2}\,\cdots$ are uncorrelated with the large scale perturbations, completely described by their own joint moments. The multiplicative constants $b_{s^2},\,b_{s^3},\,b_{\delta s^2},\,b_{st}$ and $b_\psi$ are higher order bias, and the corresponding scalars are contracted by $s^2=s_{ij}s_{ji}$, $s^3=s_{ij}s_{jk}s_{ki}$ and $st=s_{ij}t_{ji}$. In Eq.~(\ref{equ:stoch_expand}), different from the usual galaxy bias expansion, we have to further decompose the tidal field $s^2=s^2_\mathcal{D}+s^2_\mathcal{S}$, where $s^2_\mathcal{D}$ is completely correlated with matter field and resummed to the $b(k)\delta_m({\bf k})$ term, while $s^2_\mathcal{S}$ serves as one of the stochastic components with $\langle s^2_\mathcal{S}\,\delta_m\rangle=0$. The same arguments are applied to components $s^3$, $\delta_ms^2$, $st$ and $\psi$. 
Taking the Fourier transform of Eq.~(\ref{equ:stoch_expand}), we obtain
the stochasticity power spectrum,
\begin{eqnarray}
    \label{equ:stoch_pk}
    P_{\mathcal{S}}(k) &=\;& P_{\varepsilon_0\varepsilon_0}(k) + \int \frac{d^3{\bf q}}{(2\pi)^3}\, P_{\varepsilon_1\varepsilon_1}(k) P_{mm}(|{\bf q-k}|) \nonumber\\
    & +\,& \frac{13}{9}\sigma^2_{m} P_{\varepsilon_0\varepsilon_2}(k) +b_{s^2}^2 P_{s^2, \mathcal{S}}(k)  +\mathcal{O}(\varepsilon^5)  \quad ,
\end{eqnarray}
where the RMS of the fluctuation of the matter field is defined by $\sigma^2_{m} \equiv \int\,P_{mm}(q)q^2dq\,/(2\pi^2)$. 

Eq.~(\ref{equ:stoch_expand}) shows that, although $\delta_{\mathcal{S}}({\bf x})$ does not contribute to cross power spectrum, it still contributes in the higher order statistics in a complex manner.
We can classify these terms in Eq.~(\ref{equ:stoch_expand}) into three types. The first type is the isolated stochasticity parameter $\varepsilon_0$, which would contribute common shot noise $P_{\varepsilon_0\varepsilon_0} \rightarrow 1/\bar{n}_g$ for Poisson sampling. The second type is the coupling term between the stochastic variables $\varepsilon_1, \varepsilon_2, \varepsilon_{s^2}$ and the large scale fluctuation field, reflecting the coevolution of gravity and galaxy stochasticity \cite{desjacques2018large}. The coevolution also appears in the coupling terms in power spectrum Eq.~(\ref{equ:stoch_pk}), in the form of four points correlation or higher order correlation. The third type is the residual decorrelation of the underlying large scale fluctuation field with respect to the matter field, and these terms do not couple the stochastic variables in the power spectrum.

The parameterization of $r^2(k)$ is equivalent to parameterize the stochasticity power spectrum Eq.~(\ref{equ:stoch_pk}). Though some stochasticity effects have been investigated analytically, i.e. Ref.~\cite{baldauf2013halo} simplify the dark matter halo as hard sphere to derive the halo exclusion effect, it is still unable to model galaxy stochasticity in a comprehensive way due to the complex nature of galaxy formation and evolution. In order to construct an accurate parameterization for $r^2(k)$, we draw inspiration from the usual methodology in galaxy bias expansion, where the leading order stochasticity power spectrum $P_{\varepsilon_0\varepsilon_0}(k)$ is expanded as a series of $k$ and keep only the first two terms \cite{desjacques2018large, cabass2020likelihood, eggemeier2019bias, alexander2020testing, eggemeier2021testing, pezzotta2021testing, kokron2022priors, mergulhao2023effective}. Therefore, the stochasticity power spectrum is parameterized as
\begin{equation}
    \label{equ:fit_pk,1}
    P_{\mathcal{S}}(k) = \frac{\alpha}{\bar{n}_g} (1+ \beta k^2 +\gamma k^4 ) + \mathcal{O}(k^6)  \quad .
\end{equation}
\textcolor{black}{ Because of the isotropy of the stochastic field, the terms with odd power of ${\bf k}$ vanish since Fourier modes ${\bf k}$ and $-{\bf k}$ contribute equally, and consequently only the terms with even power of ${\bf k}$ remain. }
The multiplication constant $\alpha/\bar{n}_g$ accounts for the Poisson-like noise, and the $k$ series in bracket describe the scale dependence of stochasticity. As mentioned above, the previous PCA investigations have strong implication that the effective ${\rm DoF}\simeq 2$, namely two free parameters are sufficient to characterize stochasticity. Therefore, we propose the first parameterization formula in article, 
\begin{equation}
    \label{equ:fit_r2_form1}
    r^2(k) =\;\frac{1}{1+ \alpha \left(1+\beta k^2\right) /\left(\bar{n}_g P_{gg,s}(k)\right) }  \quad ,
\end{equation}
where $(\alpha, \beta)$ are free parameters and observables $P_{gg,s}$ replace $b^2P_{mm}$ compared to definition Eq.~(\ref{equ:r2}). Additionally, we keep $\alpha$ as the only non-zero free parameter to investigate the ability of 1-parameter fit describing stochasticity, and also test the 3-parameter fit to investigate the robustness of the assumption ${\rm DoF}\simeq 2$.

We can replace the measurable $P_{gg}$ with theoretical $P_{mm}$, which is determined simultaneously during the matter clustering modeling and cosmological parameters constraint (see $\texttt{Expan-mm-2}$ in Table~\ref{table:formula}), and now the parameterization is free from uncertainty in $P_{gg}$ measurement. 
Compared to Eq.~(\ref{equ:fit_r2_form1}), $\alpha$ is roughly rescaled by the square of linear galaxy bias, and the quadratic term absorbs the scale dependence of galaxy bias to some extent. In large scale limit, the absorption of the scale dependence by quadratic term is expected, since the leading order scale dependence arising from $\nabla^2\delta$ contributes to $b(k)$ scales as $k^2$ \cite{mcdonald2009clustering}.
Another straightforward thinking is whether the direct expansion of $r^2(k)$ with respect to $k$ is better than the expansion of $P_{\mathcal{S}}$. So we expand the scale dependence of $r^2$ as $k$ series and keep the first correction term, then obtain a new parameterization form (see $\texttt{Expan-new-2}$ in Table~\ref{table:formula}).

Another class of parameterization schemes inspired by the modeling of galaxy peculiar velocity in Refs.~\cite{zhang2013peculiar, zheng2013peculiar}, in which the ratio between the velocity divergence component $\theta$ and the matter overdensity $\delta_m$ is statistically compacted as a window function $W(k)$, proportional to their cross correlation coefficient $r_{\theta\delta}$. $W(k)$ describes both the non-linear evolution and the stochastic component between two fields $(\theta, \delta_m)$, similar to $(\delta_g, \delta_m)$ relation in our case. Further its fitting formula motivated by perturbation theory achieves accuracy within $2\%$ in mildly non-linear region \cite{zheng2013peculiar}. With these considerations, we adopt the proposed parameterized form of $W(k)$ for the $r^2(k)$ fit, and multiply the Poisson-like term in Eq.~(\ref{equ:fit_r2_form1}) to mimic the suppression due to the intrinsic discrete noise.
\begin{equation}
    \label{equ:fit_form_3}
    r^2(k) \;=\; \frac{1}{1+\beta \Delta_{mm}^2}\, \frac{1}{ 1+ \alpha/\left(\bar{n}_g P_{mm}\right) }  \quad .
\end{equation}
Here $\Delta_{mm}^2(k) \equiv k^3 P_{mm}(k) /(2\pi^2)$ is the matter power spectrum variance, and $(\alpha, \beta)$ are the free parameters. Similarly, we can replace $P_{mm}$ with $P_{gg}$ and obtain a new parameterization.

The most straightforward parameterization scheme is to resum the $k$- $\&$ $k^2$- dependence of $r_s^2$ to quadratic expansion $r_s^2 = 1 +c_1 k + c_2 k^2$, where $r_s\rightarrow 1$ as $k\rightarrow 0$.
Physically, we expected all the stochasticity are localized within scale $R_*$, where $R_*$ is the typical scale of galaxy formation \cite{desjacques2018large}, because galaxy distribution in very large scale is only determined by the large scale fluctuation traced by matter field. However, it is not rigorously true, because the constant discrete noise deviates from $1/\bar{n}_g$ by the sub-Poisson suppression or super-Poisson enhancement. Nevertheless, in practical cases, it does not prohibit the excellent performance of the formula such as $r_s^2 = 1 +c_1 k + c_2 k^2$. On the one hand, we expect deviation of $r_s$ from $1$ is negligible, since the clustering signal dominates in large scale and cosmic variance is significant. On the other hand, in realistic case, the formula allows $r_s^2 < 1$ at the largest scale by the adjustment of the $c_1, c_2$ terms. Therefore, we expect the quadratic formula is accurate if we set a small enough $k_{\rm max}$ during fitting. We also further reduce the number of free parameters by keeping only the quadratic terms (see $\texttt{Quadratic-1}$ in Table~\ref{table:formula}). Besides, based on the analysis of mock galaxy catalogs, the non-linear scale dependence of galaxy bias is proposed to be parameterized as $\left(1 +c_2 k^2\right)/\left(1+ c_1 k\right)$, known as 'Q-bias' prescription \cite{Smith_2007, Cole_2005, song2015consistent, bose2023euclid}. We migrate it to our stochasticity description.

The parameterization schemes are summarized in Table~\ref{table:formula}. In the following context, we use the label shown in the first column of the table to refer to these parameterization schemes. 
There are two distinctive types of parameterization for $r^2$ and $r^2_s$ when applying to realistic galaxy survey. One type of parameterization includes $P_{gg,s}$ that is directly measurable in the realistic data. While another type of parameterization includes $P_{mm}$ that is simultaneously determined in the theoretical fitting framework. We omit this distinction below for the purpose of testing.


\begin{table*}
    \renewcommand{\arraystretch}{1.3}
\caption{\label{table:galaxy_cosmoDC2} Galaxy samples: cosmoDC2 . }
\begin{ruledtabular}
\begin{tabular}{cclllll}
    \multirow{2}{*}{$\mathsf{FluxLimitLabel}$} & \multirow{2}{*}{ Flux-limit of $ugrizY$ bands } & \multicolumn{4}{c}{ $n_g\;[\,{\rm Mpc^{-3}h^3}\,]$ }  \\ 
    & &z=0.15 &z=0.5 &z=1.0 &z=1.5 \\\hline
    1 &( 28.1, 29.4, 29.5, 28.8, 28.1, 26.9 ) & 0.0845 & 0.0592 & 0.0582 & 0.0332 & \\
    2 &( 27.1, 28.4, 28.5, 27.8, 27.1, 25.9 ) & 0.0823 & 0.0494 & 0.0458 & 0.0240 & \\
    3 &( 26.1, 27.4, 27.5, 26.8, 26.1, 24.9 ) & 0.0793 & 0.0394 & 0.0344 & 0.0167 & \\
    4 &( 25.1, 26.4, 26.5, 25.8, 25.1, 23.9 ) & 0.0749 & 0.0300 & 0.0256 & 0.0111 & \\
    5 &( 24.1, 25.4, 25.5, 24.8, 24.1, 22.9 ) & 0.0686 & 0.0225 & 0.0196 & 0.0067 & \\
    6 &( 23.1, 24.4, 24.5, 23.8, 23.1, 21.9 ) & 0.0609 & 0.0163 & 0.0140 & 0.0033 & \\
    7 &( 22.1, 23.4, 23.5, 22.8, 22.1, 20.9 ) & 0.0520 & 0.0115 & 0.0090 & 0.0014 & \\
    8 &( 21.1, 22.4, 22.5, 21.8, 21.1, 19.9 ) & 0.0428 & 0.0076 & 0.0057 & 0.0005 & \\
\end{tabular}
\end{ruledtabular}

\caption{\label{table:galaxy_TNG300} Galaxy samples: TNG300-1 . }
\begin{ruledtabular}
\begin{tabular}{cclllll}
    \multirow{2}{*}{$\mathsf{FluxLimitLabel}$} & \multirow{2}{*}{ Flux-limit of $griz$ bands } & \multicolumn{4}{c}{ $n_g\;[\,{\rm Mpc^{-3}h^3}\,]$ }  \\ 
    & &z=0.15 &z=0.5 &z=1.0 &z=1.5 \\\hline
    1 &( 29.4, 29.5, 28.8, 28.1) & 0.1353 & 0.0915 & 0.0698 & 0.0550 & \\
    2 &( 28.4, 28.5, 27.8, 27.1) & 0.1221 & 0.0794 & 0.0587 & 0.0470 & \\
    3 &( 27.4, 27.5, 26.8, 26.1) & 0.1129 & 0.0700 & 0.0497 & 0.0377 & \\
    4 &( 26.4, 26.5, 25.8, 25.1) & 0.1032 & 0.0611 & 0.0422 & 0.0298 & \\
    5 &( 25.4, 25.5, 24.8, 24.1) & 0.0914 & 0.0525 & 0.0352 & 0.0244 & \\
    6 &( 24.4, 24.5, 23.8, 23.1) & 0.0815 & 0.0439 & 0.0295 & 0.0198 & \\
    7 &( 23.4, 23.5, 22.8, 22.1) & 0.0707 & 0.0379 & 0.0239 & 0.0163 & \\
    8 &( 22.4, 22.5, 21.8, 21.1) & 0.0608 & 0.0316 & 0.0195 & 0.0125 & \\
\end{tabular}
\end{ruledtabular}
\end{table*}


\section{Methodology \label{sec:methodology}}

We present the accuracy analysis of the parameterization schemes in two ways: 
(1) Quantify absolute difference between the measurable $\hat{r}$ and the fit values $r$ by quantity ${\bf Q}$.
(2) Investigate the impact on the matter clustering amplitude $A$ by the parameterization schemes, where the accuracy and precision of $A$ is quantified by the systematic bias $\delta_A$ and statistic uncertainty $\sigma_A$.

We measure $\hat{r}^2$ in the simulations, and then fit with the parameterized formula $r^2$ by minimizing $\chi^2$.
\begin{eqnarray}
    \label{equ:chi2_fitting}
    \chi^2 = \sum_{k=k_{\rm min}}^{k_{\rm max}}{ \frac{ \left[ \hat{r}^2(k) - r^2(k) \right]^2 }{\sigma^2(k)} }  \quad ,
\end{eqnarray}
where $\sigma^2$ is approximated as the Gaussian variance of the measured $r^2$ or $r^2_s$.
In principle, both the matter and galaxy overdensity field are non-Gaussian, especially in the small scale region where the strong gravitational clustering leads to non-linear evolution. The Gaussian variance only accounts for part of the fluctuation in overdensity field, and the underestimation of Gaussian variance increase severely as deep into non-linear scale. So $\chi^2$ defined by Eq.~(\ref{equ:chi2_fitting}) does not accurately characterize the statistic significance of the parameterization fitting with measurement. However, it still serves as a practical criterion to fit the data and find the bestfit parameters. To avoid the computational challenges of estimating large amount of covariance of test samples, we utilize the absolute difference of $r_{(s)}$ to quantify the goodness of fit, rather than the $\chi^2$ statistic.
We define the absolute deviation between the measured $\hat{r}$ and $r$ over the fit range $(k_{\rm min},\,k_{\rm max})$
\begin{eqnarray}
    \label{equ:Qval_define}
    {\bf Q} \;=\; \sqrt{ \frac{ \sum_{i}{ N_{i}\,\left(\hat{r}_i - r_i \right)^2 } }{ \sum_i N_{i} } } \quad ,
\end{eqnarray}
here $N_i \propto k^2_i\Delta k_i$ is the number of independent $k$ modes in $i$-th $k$ bin, and ${\bf Q}$ is the RMS over all the fitting bin weighted by the number of $k$ modes in each bin.

In the cosmological applications, we are interested in the impact on the accuracy of the reconstructed matter power spectrum, rather than the goodness of fit. Specifically, we focus on two questions: how much the bias in the constrained total amplitude of the matter power spectrum (or equivalently $\sigma_8$ in linear scale) is and how much the constraint precision should be sacrificed using these parameterizations. As an illustration, we forecast these impacts by combining the projected observable $\hat{C}^{g\kappa}_\ell$ and $\hat{C}^{gg}(k)$, where the galaxy is the spectroscopic tracer that targeted by DESI \cite{abareshi2022overview, adame2023validation}, and the lensing convergence $\kappa$ could be realized by the future CSST cosmic shear survey \cite{cao2018testing, gong2019cosmology, lin2022forecast}. 

Under the Limber's approximation, the measured angular power spectrums of galaxy clustering and galaxy lensing are given by
\begin{eqnarray}
    \label{equ:C_gkappa_measured}
    \hat{C}^{g\kappa}_\ell &=& \int{ d\chi \,\chi^{-2}\, n_g(\chi) W_L(\chi) \hat{P}_{gm}\left( \frac{\ell}{\chi}, \chi \right) }  \\
    \label{equ:C_g_measured}
    \hat{C}^{gg}_\ell &=& \int{ d\chi \,\chi^{-2}\, n_g(\chi)n_g(\chi) \hat{P}_{gg}\left( \frac{\ell}{\chi}, \chi \right) }   \quad . \nonumber
\end{eqnarray}
Here $W_L(\chi)$ is the lensing kernel. 
For simplicity, we assume the both the redshift distributions of the foreground galaxy are narrow enough that \textcolor{black}{ $n_g(\chi) \rightarrow \delta^{D}(\chi-\chi_g)$ }, then we have
\begin{eqnarray}
    \label{equ:Cgkappa_simplied}
    \hat{C}^{g\kappa}_\ell  &=& \chi_g^{-2}\,W_L(\chi_g)\, \hat{P}_{gm}\left( \frac{\ell}{\chi_g}\right)  \quad ,  \\
    \label{equ:Cgg_simplied}
    \hat{C}^{gg}_\ell  &=& \chi_g^{-2}\,\Delta\chi_g^{-1}\, \hat{P}_{gg}\left( \frac{\ell}{\chi_g}\right)  \quad . 
\end{eqnarray}
Here $\Delta\chi_g$ is the comoving radical distance width of the galaxy bin. 
{ \color{black} More cautious validation and treatment of the approximation are required in realistic applications, while its sub-percent level of inaccuracy has a negligible impact on the estimation of parameterization performance. } 
The reconstructed power spectrum $\hat{P}(k)$ differs from the direct measurable $\hat{C}_\ell$ with a geometric factor. We can rewrite $\hat{P}_{gg}$ in Eq.~(\ref{equ:Cgg_simplied}) in the form of cross power spectrum.
\begin{eqnarray}
    \label{equ:C_gkappa_themplate}
    C^{g\kappa}_\ell &=& \chi_g^{-2}\,W_L(\chi_g)\, r\, \sqrt{ \hat{P}_{gg}\left( \frac{\ell}{\chi_g}\right) P_{mm}\left( \frac{\ell}{\chi_g}\right) }
\end{eqnarray}
Here $r$ is the model of cross correlation. If we subtract the Poisson expectation in the galaxy power spectrum, a replacement should be made $(r\,,\hat{P}_{gg}) \rightarrow (r_s\,,\hat{P}_{gg,s})$. Since the fluctuation in cosmic shear is the dominant noise source in the realistic measurements, we can neglect the uncertainty in $\hat{P}_{gg}$ and consider only the noise contribution from $\hat{C}^{g\kappa}_\ell$. Therefore we can treat $C^{g\kappa}_\ell$ in Eq.~(\ref{equ:C_gkappa_themplate}) as a theoretical template for the observable of Eq.~(\ref{equ:Cgkappa_simplied}). The matter power spectrum $P_{mm}$ can be estimated by minimizing 
\begin{eqnarray}
    \label{equ:chi_C_gkappa}
    \chi^2_{A} = \left[\hat{C}^{g\kappa}_\ell- C^{g\kappa}_\ell \right]^{T} {\bf Cov}^{-1} \left[\hat{C}^{g\kappa}_\ell- C^{g\kappa}_\ell \right]  \; .
\end{eqnarray}
Further, we only consider the Gaussian uncertainty, which leads to null off-diagonal elements of covariance ${\bf Cov}$ and only its diagonal elements contribute. Under these assumptions, the loss function Eq.~(\ref{equ:chi_C_gkappa}) combined with Eq.~(\ref{equ:Cgkappa_simplied}) and Eq.~(\ref{equ:C_gkappa_themplate}) can be simplified as
\begin{eqnarray}
    \label{equ:chi_biaedA}
    \chi_{A}^2 = \left[\frac{W_L(\chi_g)}{\chi_g^2}\right]^2 \sum_{\ell} \frac{ \hat{P}_{gg}P_{mm} }{ \sigma_\ell^2 } \left( \hat{r} - A\,r \right)^2
\end{eqnarray}
where we make replacement $P_{mm}\rightarrow A^2P_{mm}$ to fix $P_{mm}$ as the fiducial true value. The free rescaling factor $A$ marginalizes the overall amplitude of the measured matter field, $A=1$ meaning the unbiased measurement. $\hat{r}\equiv\hat{P}_{gm}/\sqrt{\hat{P}_{gg}P_{mm}}$ is the fiducial true cross correlation coefficient by definition. Now all the gained cosmological information are contained in the amplitude $A$, of which the systematic bias is characterized by $\delta_A = A-1$ and the statistic uncertainty is $\sigma_A$. For a stochasticity parameterization scheme given, the matter clustering amplitude as well as the nuisance stochasticity parameters $\lambda_\mu = \left(\,A, \alpha, \beta, \cdots\right)$ are given by the bestfit values of Eq.~(\ref{equ:chi_biaedA}), then the systematic bias $\delta_A$ is also determined. The statistic uncertainty $\sigma_A$ is given by the Fisher analysis \cite{schmittfull2018parameter},
\begin{eqnarray}
    \sigma^2_{\lambda_\mu} \;=\; \left( \frac{1}{2} \frac{\partial^2 \chi_A^2 }{\partial\lambda_\alpha\partial\lambda_\beta}\biggl|_{\rm bestfit} \right)^{-1} \biggl|_{\mu\mu}
\end{eqnarray}
We have assumed the Gaussian variance for measurement in Eq.~(\ref{equ:chi_biaedA}), which is given by
\begin{eqnarray}
    \label{equ:gaussion_error_chiA}
    \sigma^2_\ell &=& \frac{1}{f_{\rm sky}(2\ell+1)\Delta\ell} \left[\hat{C}^{gg}_\ell \hat{C}^{\kappa\kappa}_\ell +(C^{g\kappa}_\ell)^2 \right]  \quad .
\end{eqnarray}
Specifically, 
\begin{eqnarray}
    \hat{C}^{gg}_\ell &=& {C}^{gg}_\ell + \frac{1}{\Sigma_g}  \;, \nonumber\\
    \hat{C}^{\kappa\kappa}_\ell  &=& {C}^{\kappa\kappa}_\ell + \frac{\sigma_\gamma^2}{\Sigma_\gamma} +N^{\gamma}_{\rm add}  \;, 
\end{eqnarray}
where, $\Sigma$ is the surface number density of tracers, $\sigma_\gamma$ is the shape noise and $N_{\rm add}^{\gamma}$ is the additive noise term, all of which vary with survey details.

If the impact of $r^2$ is neglected, namely setting a model $r=1$, the systematic bias and the statistic uncertainty can be analytically expressed as
\begin{eqnarray}
    \label{equ:paraA_bias_not_r2}
    \delta_A &\,=\,& \frac{ \sum_{\ell}{\sigma_\ell^{-2}\, \hat{P}_{gg}P_{mm} \hat{r}} }{ \sum_{\ell}{\sigma_\ell^{-2}\, \hat{P}_{gg}P_{mm}} } - 1   \\
    \sigma_A &\,=\,& \left[\frac{W_L(\chi_g)}{\chi_g^2}\right]^{-1} \left(\sum_{\ell} \frac{ \hat{P}_{gg}P_{mm} }{ \sigma_\ell^2 } \right)^{-1/2} \nonumber 
\end{eqnarray}


\begin{figure*}
\includegraphics[width=1.0\textwidth]{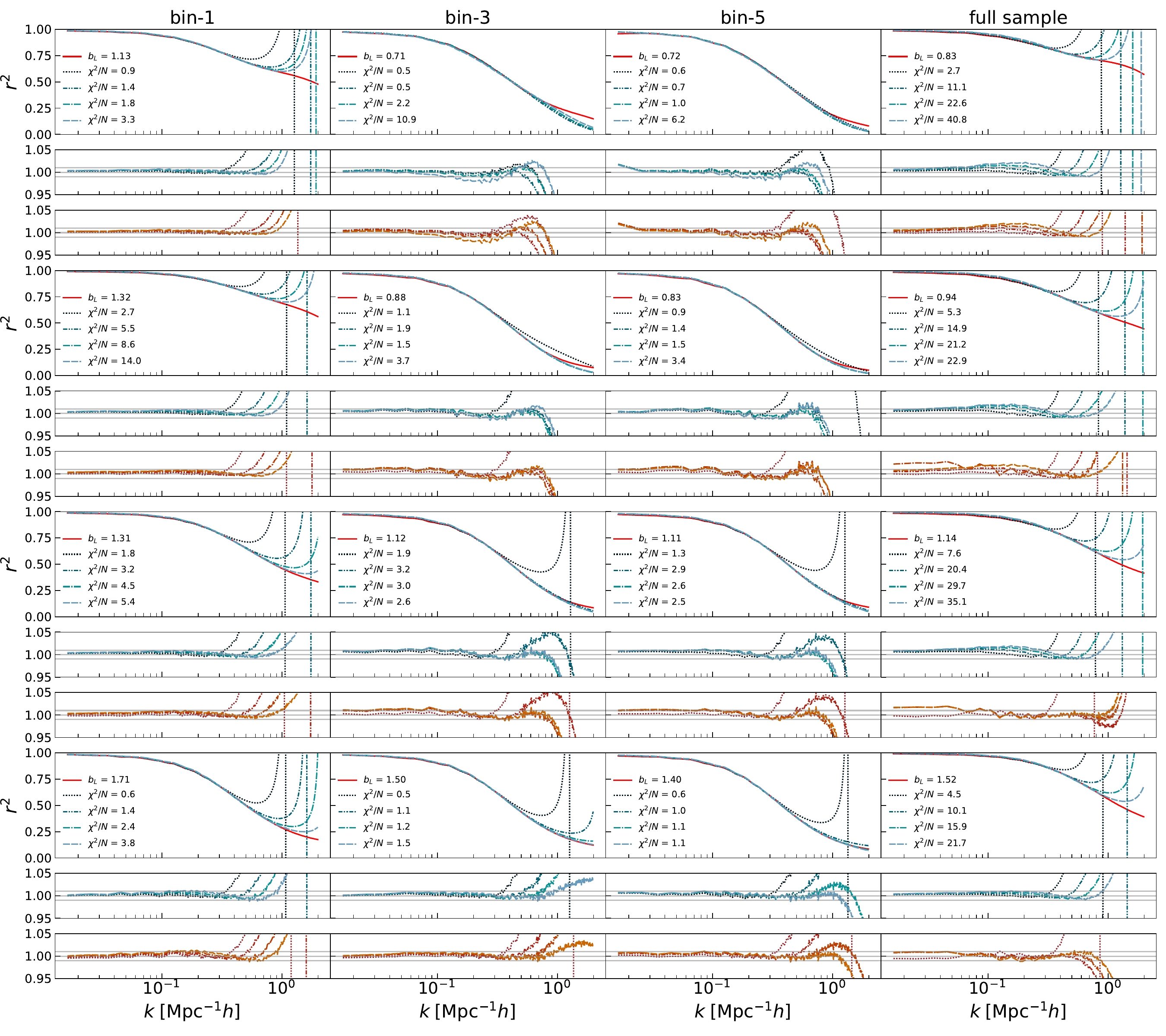}
\caption{ \label{fig:fitting_ratio_r2} The measured $\hat{r}^2$ and their fitting results $r^2$ by $\texttt{Expan-mm-2}$ and $\texttt{Zheng2013-mm}$ on cosmoDC2 galaxies ($\mathsf{FluxLimitLabel}=3$). Four groups of rows show the results at $z = 0.15, 0.5, 1.0$ and $1.5$. The first three columns show the results of 3 selected galaxy $g-r$ color bins (total 6 bins), and the last column shows the result of flux limited full samples. \textit{Thick Rows:} The solid line is $\hat{r}^2$ measured in simulation, and the dash or dot lines are the result of fitting formula $\texttt{Expan-mm-2}$ fitted with $k_{\rm max}=0.3, 0.5, 0.7$ and $0.9\,{\rm Mpc^{-1}}h$. \textit{Narrow Rows:} They show the ratio between $r^2$ and $\hat{r}^2$, and the blue lines are fitted with $\texttt{Expan-mm-2}$ while the red lines are fitted with $\texttt{Zheng2013-mm}$, where horizontal lines mark the region of $\pm 1\%$. 
Notice that $\chi^2/N$ do not serve as a goodness of fit criterion here.
}
\end{figure*}

\begin{figure*}
\includegraphics[width=1.0\textwidth]{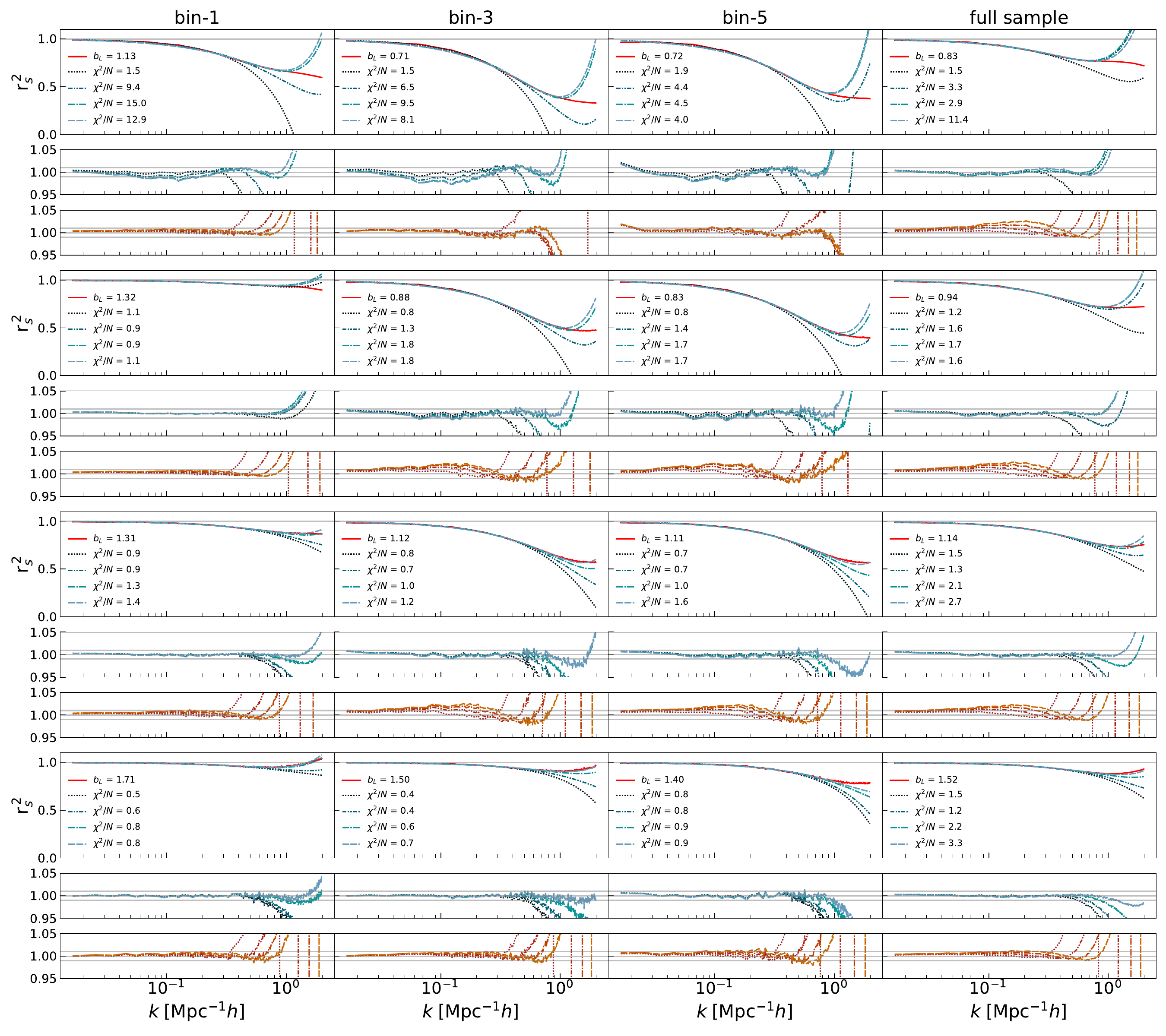}
\caption{ \label{fig:fitting_ratio_r2s} Same plot as Fig.~\ref{fig:fitting_ratio_r2} for cosmoDC2 galaxies ($\mathsf{FluxLimitLabel}=3$), but here we present results for $r^2_s$. The blue lines are fitted with $\texttt{Quadratic}$ while the red lines are fitted with $\texttt{Expan-mm-2}\; (r_s^2)$.  }
\end{figure*}


\begin{table*}
    \renewcommand{\arraystretch}{1.3}
\caption{\label{table:SummarizeIndex} 
Summarized Index for Stochasticity Parameterization. Two values provided for each index shown in form $A\,|\,B$, meaning the values measured in cosmoDC2 ($A$) or TNG300-1 ($B$) galaxy samples. The range of measurement is $0.03 <k< 0.9\;{\rm Mpc^{-1}h}$. The bar over the quantity means the value averaging over the all the fiducial samples of 4 redshift snapshots, 8 $\mathsf{FluxLimitLabel}$, 15$\,|\,$6 galaxy color and 6 bins for a color bin. 
}
\begin{ruledtabular}
\begin{tabular}{cccccccc}
    & &  $\bar{{\bf Q}}\times 100$ 
    & worst ${\bf Q}\times 100$ & $\bar{\delta}_A$  & $\bar{\sigma}_A$ & worst $\delta_{A}$ & \\ \hline
    & $\texttt{Expan-mm-1}$ & $1.92\,|\,1.41$ & $4.65\,|\,3.92$ & $0.095\,|\,0.044$ & $0.040\,|\,0.031$ & $0.327\,|\,0.112$ &   \\
    & $\texttt{Expan-mm-2}$ & $0.83\,|\,0.71$ & $1.45\,|\,1.27$ & $0.019\,|\,0.013$ & $0.043\,|\,0.037$ & $0.133\,|\,0.048$ &   \\
    & $\texttt{Expan-mm-3}$ & $0.77\,|\,0.65$ & $1.05\,|\,0.95$ & $0.011\,|\,0.010$ & $0.049\,|\,0.043$ & $0.071\,|\,0.029$ &   \\
    & $\texttt{Expan-gg-1}$ & $1.07\,|\,0.86$ & $2.62\,|\,1.63$ & $0.048\,|\,0.016$ & $0.035\,|\,0.028$ & $0.301\,|\,0.042$ &   \\
    & $\texttt{Expan-gg-2}$ & $0.87\,|\,0.71$ & $1.05\,|\,1.03$ & $0.013\,|\,0.012$ & $0.040\,|\,0.035$ & $0.094\,|\,0.026$ &   \\
    & $\texttt{Expan-gg-3}$ & $0.87\,|\,0.69$ & $1.06\,|\,1.04$ & $0.010\,|\,0.010$ & $0.047\,|\,0.041$ & $0.050\,|\,0.021$ &   \\
    & $\texttt{Expan-new-2}$ & $0.92\,|\,0.72$ & $1.65\,|\,1.20$ & $0.025\,|\,0.011$ & $0.043\,|\,0.037$ & $0.103\,|\,0.031$ &   \\
    & $\texttt{Zheng2013-mm}$ & $0.72\,|\,0.70$ & $1.59\,|\,2.08$ & $0.015\,|\,0.009$ & $0.042\,|\,0.037$ & $0.116\,|\,0.033$ &   \\
    & $\texttt{Zheng2013-gg}$ & $0.79\,|\,0.72$ & $1.65\,|\,1.80$ & $0.012\,|\,0.007$ & $0.042\,|\,0.038$ & $0.106\,|\,0.016$ &   \\    \cline{2-7}
    & $\texttt{Expan-mm-1 ($r_s^2$)}$ & $1.03\,|\,0.85$ & $3.38\,|\,2.00$ & $0.020\,|\,0.021$ & $0.030\,|\,0.027$ & $0.039\,|\,0.044$ &   \\
    & $\texttt{Expan-mm-2 ($r_s^2$)}$ & $0.65\,|\,0.51$ & $3.14\,|\,0.87$ & $0.011\,|\,0.014$ & $0.038\,|\,0.034$ & $0.023\,|\,0.030$ &   \\
    & $\texttt{Expan-new-2 ($r_s^2$)}$ & $0.73\,|\,0.57$ & $3.11\,|\,0.98$ & $0.010\,|\,0.012$ & $0.040\,|\,0.036$ & $0.019\,|\,0.026$ &   \\
    & $\texttt{Zheng2013-mm ($r_s^2$)}$ & $1.07\,|\,0.93$ & $3.51\,|\,1.95$ & $0.008\,|\,0.011$ & $0.039\,|\,0.035$ & $0.019\,|\,0.030$ &   \\
    & $\texttt{Q-bias}$ & $0.55\,|\,0.26$ & $3.15\,|\,0.86$ & $0.018\,|\,0.012$ & $0.050\,|\,0.045$ & $0.051\,|\,0.034$ &   \\
    & $\texttt{Quadratic}$ & $0.35\,|\,0.19$ & $3.15\,|\,0.42$ & $0.006\,|\,0.006$ & $0.037\,|\,0.037$ & $0.024\,|\,0.018$ &   \\
    & $\texttt{Quadratic-1}$ & $5.41\,|\,3.62$ & $14.86\,|\,12.37$ & $0.088\,|\,0.070$ & $0.016\,|\,0.015$ & $0.185\,|\,0.166$ &   \\
\end{tabular}
\end{ruledtabular}
\end{table*}


\begin{figure*}
\includegraphics[width=0.98\textwidth]{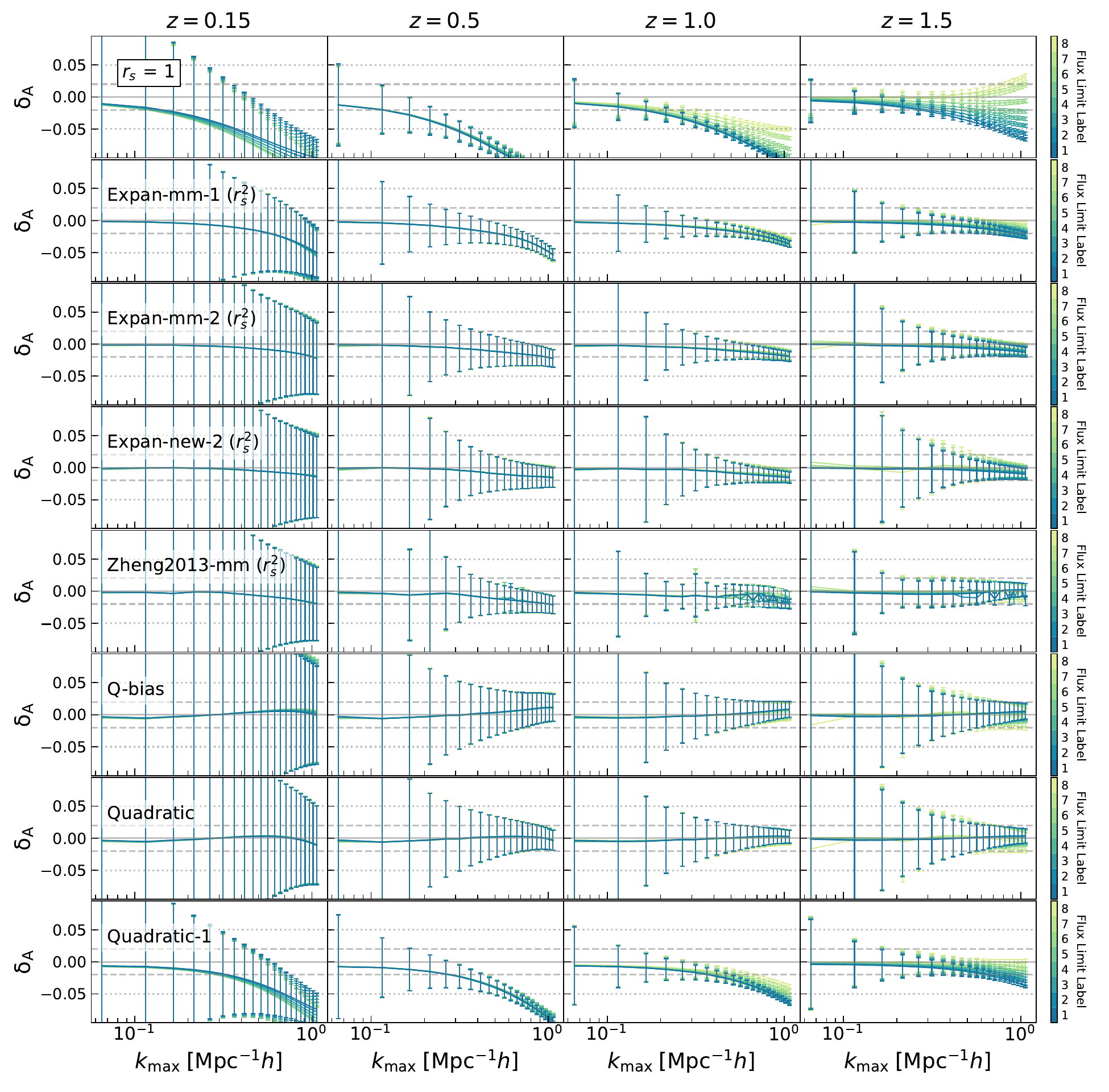}
\caption{ \label{fig:biasA_cosmoDC2-1} The matter clustering amplitude $A$ measured in cosmoDC2 galaxies. We show the bias $\delta_A=A-1$ and corresponding uncertainty $\sigma_A$ as function of maximum fit scale $k_{\rm max}$. Four rows show the results at redshift $z = 0.15, 0.5, 1.0$ and $1.5$ respectively. The first column shows the results setting $r_s=1$, and the rest of 7 columns show the results for $r^2_s$ parameterizations.
In each subfigure, different colors denote the full flux limited samples with different $\mathsf{FluxLimitLabel}$, and the dash/dot lines mark the region of $\pm 2\%\,/\,\pm 5\%$. }
\end{figure*}

\begin{figure*}
\includegraphics[width=0.98\textwidth]{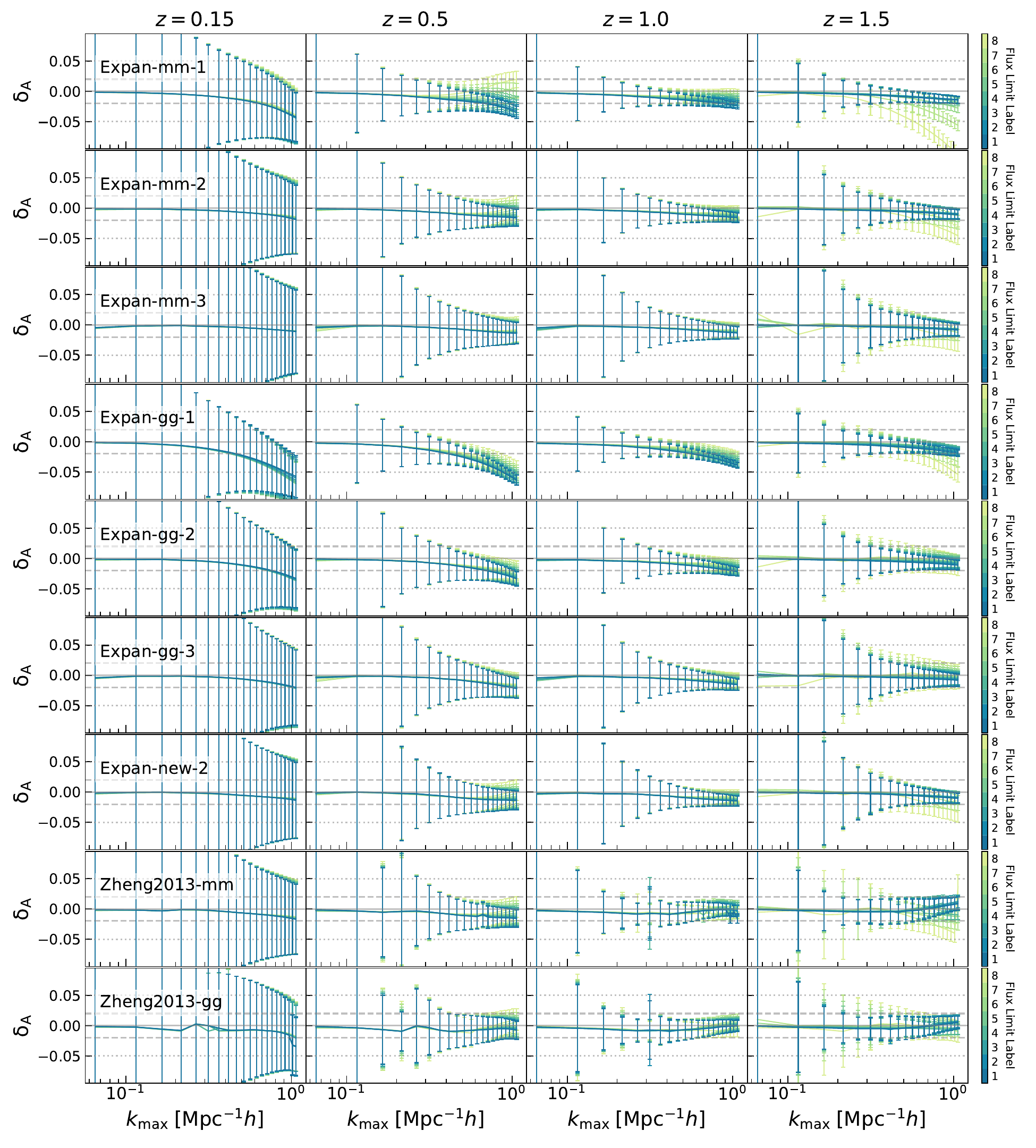}
\caption{ \label{fig:biasA_cosmoDC2-2} Same plot as Fig.~\ref{fig:biasA_cosmoDC2-1} for the full flux limited samples with different $\mathsf{FluxLimitLabel}$ in cosmoDC2 galaxies. Here we present results for $r^2$ parameterizations.  }
\end{figure*}


\begin{figure}
\includegraphics[width=1.0\columnwidth]{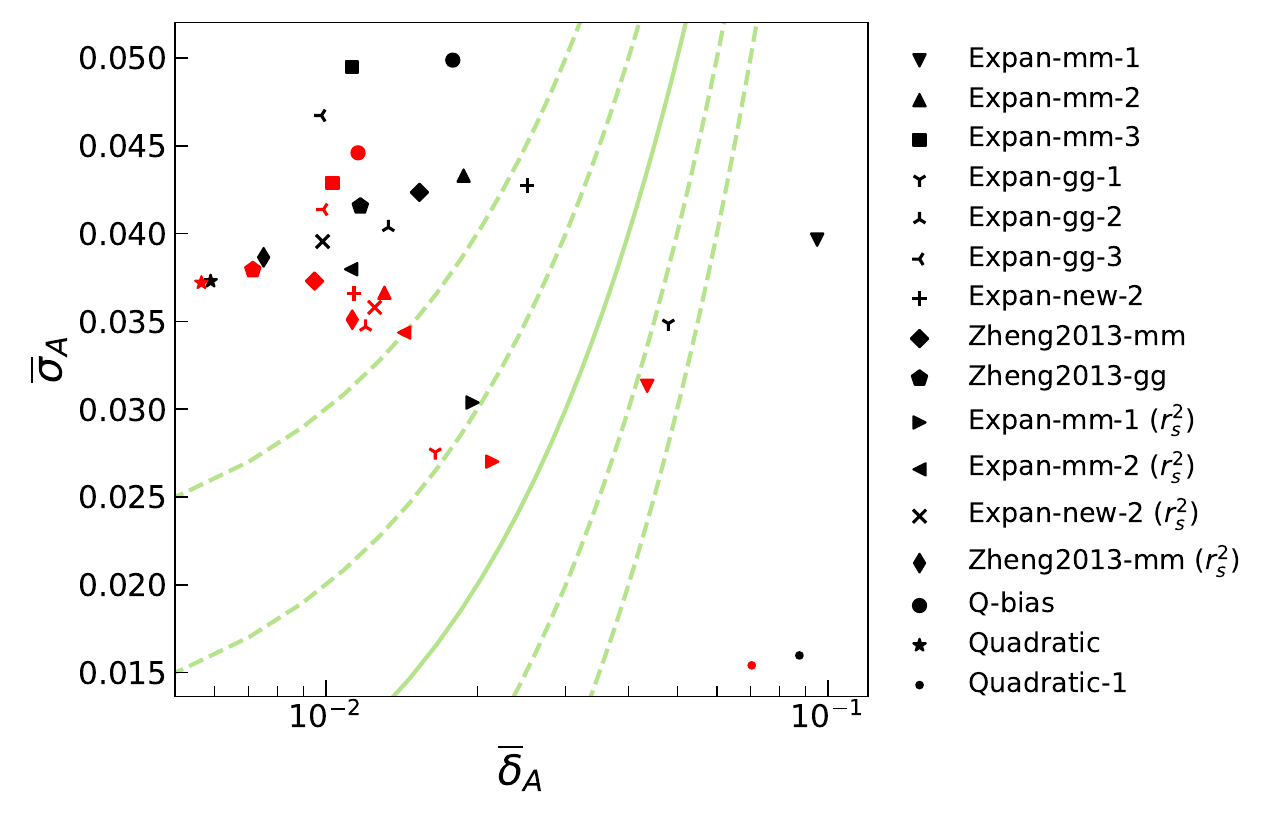}
\caption{ \label{fig:summarizedIndex} Scatter plot of relation between $\bar{\delta}_A$ and $\bar{\sigma}_A$ of fourth and fifth columns in Table~\ref{table:SummarizeIndex}. Black/Red markers denote the results of cosmoDC2/TNG300-1 galaxies. The green solid line marks $\bar\sigma_A=\bar\delta_A$, and the dash lines mark $\bar\sigma_A=\bar\delta_A\pm1\%$ and $\bar\sigma_A=\bar\delta_A\pm2\%$. \textcolor{black}{Notice that we use the logarithm scaling in x-axis for distinguishing the scatter points. }
}
\end{figure}


\section{ Performance of Proposed Parameterizations \label{sec:results}}

In this section, we present the results of testing the fitting accuracy and reconstruction ability of the parameterizations. In Section~\ref{subsec:result_1}, we present the direct $r^2$ fitting accuracy for our fiducial galaxy samples. In Section~\ref{subsec:result_2}, we further analyze the impact on the reconstructed amplitude $A$, based on the fiducial galaxy samples and assuming the cosmic variance limit for lensing measurement.

The fiducial galaxy samples for testing stochasticity consist of two sets of galaxy mocks, cosmoDC2 galaxy mock \cite{korytov2019cosmodc2} and TNG300-1 galaxy samples \cite{Springel_2017, Nelson_2017, Pillepich_2017, Naiman_2018, Marinacci_2018}, combined with $1200\,h^{-1}{\rm Mpc}$ dark matter simulation. The details of fiducial galaxy samples are listed in Table~\ref{table:galaxy_cosmoDC2} $\&$ \ref{table:galaxy_TNG300}. 
The galaxy samples are defined by the redshift, flux and the galaxy color. We apply various magnitude cut of $ugrizY$/$griz$ bands to the cosmoDC2/TNG300-1 galaxies. These flux limited samples are labeled by tag $\mathsf{FluxLimitLabel}$. Notice that $\mathsf{FluxLimitLabel}=3$ corresponds to the expected depth of coadded images from LSST \cite{ivezic2019lsst}. In order to test our parameterization with various stochasticity, we further separate the full flux limited galaxy sample into multi-bins according to the galaxy color. For cosmoDC2 galaxy, we utilize all the possible color by $ugrizY$ bands combination ($u-g$, $g-r$ and so on), obtaining totally $15$ kinds of galaxy colors, and for TNG300-1 there is $6$ kinds of galaxy colors. The number of galaxy color bins is $N_{\rm bins}=6$, and the number of galaxies in each bin is identical. We also get the consistent results with $N_{\rm bins}=12$, but we do not show details here to avoid redundancy.


\subsection{Goodness of Fit \label{subsec:result_1}}

In Fig.~\ref{fig:fitting_ratio_r2}, we show the measured $\hat{r}^2$ on cosmoDC2 galaxy samples and their fitting results $r^2$ by schemes $\texttt{Expan-mm-2}$ and $\texttt{Zheng2013-mm}$. As the fit range $k_{\rm max}\lesssim 0.5 \;{\rm Mpc^{-1}h}$, the accuracy generally  reaches $\lesssim1\%$. When $k_{\rm max}$ extends to $0.9 \;{\rm Mpc}^{-1}h$, it still maintains the accuracy level at $3\%$ conservatively, and for most of the samples shown in figure, we still have accuracy $\lesssim 1\%$. 
In Fig.~\ref{fig:fitting_ratio_r2s}, we show the same results for $r^2_s$ by $\texttt{Quadratic}$ and $\texttt{Expan-mm-2}$. We can see a tighter fitting for $r_s^2$, especially for the full sample, the $\texttt{Quadratic}$ accuracy is strictly at the level of $< 1\%$ at $k_{\rm max}<0.9{\,\rm Mpc^{-1}}h$ for samples at redshift $z>0.5$. A visual advantage of $r_s^2$ fit results is that, our parameterizations are able to characterize the upwarp around $k\sim 1{\,\rm Mpc^{-1}}h$ due to the halo exclusion effect, for instance $\texttt{Quadratic}$ shown in Fig.~\ref{fig:fitting_ratio_r2s}. The terms such as $k^2$ are unfavorable in direct $P_{\mathcal{S}}$ fitting, due to the divergence in Fourier transform, but fortunately this property is able to capture the sub-Poisson behavior appearing in the non-linear scale, enhancing the flexibility of $r_s^2$ fitting in turn.
Similar results are obtained for other parameterization schemes.

We utilize the mean ${\bf\bar{Q}}$ values to summarize the performance of the proposed parameterizations against the above galaxy samples. Here, ${\bf\bar{Q}}$ means averaging ${\bf Q}$ values over the all the galaxy samples of 4 redshift snapshots, 8 $\mathsf{FluxLimitLabel}$, 15 (6) galaxy color and 6 bins for a color bin. The range of k is fixed as $0.03\;{\rm Mpc^{-1}}h <k< 0.9\;{\rm Mpc^{-1}}h$, where the larger scale modes $k\lesssim 0.03{\,\rm Mpc^{-1}}h$ are affected greatly by cosmic variance and the definition of $\bf Q$ does not account for the fluctuation in measurement. We list the results in first and second columns in Table~\ref{table:SummarizeIndex}.
Based on ${\bf\bar{Q}}$ comparison, we can derive two immediate conclusions:
\begin{itemize}
\item[$\bf (1)$] Two free parameters in parameterization are able to describe stochasticity to small scale $k_{\rm max}= 0.9\;{\rm Mpc^{-1}}h$ in $\sim1\%$ accuracy, verifying the implication by the previously found ${\rm DOF}\simeq 2$ eigenmodes in simulated galaxy clustering.
\item[$\bf (2)$] The $\texttt{Quadratic}$ scheme is the most promising parameterization with the best $<0.5\%$ accuracy in $r_s^2$ fitting among all 12 proposed parameterization schemes, quantified by the summarized performance over all the samples.
\end{itemize}
Specifically, the inclusion of more free parameters do benefit the $r^2_{(s)}$ fitting. For instance, when we replace the scheme $\texttt{Expan-mm-1}$ by $\texttt{Expan-mm-2}$, the ${\bf\bar{Q}}$ value decreases $1.92\%\rightarrow 0.83\%$ for cosmoDC2 galaxies and $1.41\%\rightarrow 0.71\%$ for TNG300-1 galaxies, almost twice as much. However, the ${\bf\bar{Q}}$ decrement when replacing $\texttt{Expan-mm-2}$ by $\texttt{Expan-mm-3}$ is not as much as for $\texttt{Expan-mm-1}$ $\rightarrow$ $\texttt{Expan-mm-2}$, producing little improvement especially for $\texttt{Expan-gg-2}$ $\rightarrow$ $\texttt{Expan-gg-3}$. 
For all $2$-parameter $r^2$ parameterizations, we have ${\bf\bar{Q}}\sim 0.8\%$, and the worst values appears among all the galaxy samples is ${\bf Q}< 2\%$. For all $2$-parameter $r^2_s$ parameterizations, we have similar value ${\bf\bar{Q}}\lesssim 1\%$ for all samples, but it seems that worst ${\bf Q}\sim 3.5\%$ of cosmoDC2 samples is larger than that for $r^2$. It arises from the intrinsic true that $r_s > r$, and the deviation $\delta r$ is estimated by ${\bf Q}(r)\sim \delta r$. Therefore we expect that ${\bf Q}(r_s) > {\bf Q}(r)$ in general, and the larger ${\bf Q}(r_s)$ value compared to ${\bf Q}(r)$ do not confirm $r^2$ fits outperforms $r^2_s$ fits. Even so, some $r^2_s$ parameterizations still achieve a smaller ${\bf\bar{Q}}$ value compared to $r^2$ parameterizations, especially $\texttt{Quadratic}$, of which the ${\bf\bar{Q}}$ is smaller than any other parameterization even those with $3$ parameters. It suggests that $\texttt{Quadratic}$ is a promising candidate to characterize the stochasticity.

Besides the fiducial color-bins samples, we further separate the full flux limited galaxy samples into multi-bins by stellar mass and host-halo mass. In the stellar mass classified galaxy samples, the parameterization performance is consistent with the fiducial color-bins case. The absolute values of the measured $\bf \bar{Q}$ shows little difference compared to the color-bins samples, and the $\texttt{Quadratic}$ scheme is still the best performing parameterization with ${\bf\bar{Q}}=0.39\%\,\&\,0.17\%$ for cosmoDC2$\,\&\,$TNG300-1 samples, closed to the fiducial case. However, the situation changes if we adopt host-halo mass as the classification criteria, since it separates the galaxies into satellite-rich and satellite-poor galaxy populations, as well as into the halo exclusion-effective and exclusion-ineffective galaxy populations. As it has derived in \cite{baldauf2013halo,kokron2022priors}, the large scale modes can not identify the galaxies in a same host-halo, resulting in the host-halo dominating shot noise in large scale, sourcing the sup-Poisson enhancement. So the satellite-poor bin is affected less by the enhancement, while the satellite-rich bin is enhanced greatly. Apart from the enhancement, the halo exclusion effect is more and more significant as the host-halo mass increase, and it sources the sub-Poisson suppression. 
Within our performance tests on cosmoDC2$\,\&\,$TNG300-1, we find the sup-Poisson enhancement is dominant in the first 5 less-massive bins, and the there is a trend of increasing $r_s^2$ suppression as host-halo mass increases. For these 5 bins, a particular feature is the significant suppression $r_s^2\lesssim 0.95$ even just at $k=0.05{\,\rm Mpc^{-1}}h$, distinguished from the color-bins as well as the stellar mass-bins samples. While for the most massive bin, the halo exclusion overcomes the sup-Poisson effect and alleviates the suppression on $r_s^2$, allowing $r_s^2\sim 1$ for cosmoDC2 galaxies. 
Despite significant distinction in these host-halo mass bin samples, our 2- $\&$ 3-parameter parameterizations still work well within range $0.03<k<0.9{\,\rm Mpc^{-1}}h$. In contrast, the 1-parameter fits such as $\texttt{Expan-mm-1}$ reach ${\bf\bar{Q}}=3.7\%\,\&\,3.6\%$, revealing that one degree of freedom is not flexible enough to parameterize the stochasticity. For the 2-parameter fitting, ${\bf\bar{Q}}$ deviates slightly from the results in Table~\ref{table:SummarizeIndex}, and the worst $\bf Q$ appearing in $r^2$ parameterizations is limited to ${\bf Q}<3\%$. Especially, $\texttt{Quadratic}$ still provides the best performance with ${\bf\bar{Q}}=0.53\%\,\&\,0.61\%$. To sum up, all the tests on various samples reinforce two conclusions above, reiterating the suggestion that the simple $\texttt{Quadratic}$ is a promising choice for stochasticity parameterization


\subsection{Impact on Matter Clustering Amplitude \label{subsec:result_2}}

We utilize $\delta_A$ and $\sigma_A$ to quantify the ability of parameterizations alleviating the stochasticity impact. Solely for the purpose of comparison, we assume the distribution of the lensed galaxy sources is an infinite thin plane, locating at $z_s = 3$, and the redshift bin width of the foreground clustering galaxies is $\Delta z = 0.1$, with $f_{\rm sky}=1$ for both sets. For simplicity, we assume cosmic variance limit for the cosmic shear measurements, equivalently neglecting shotnoise in $\kappa$ field. The neglected complexities such as the redshift distribution of source galaxies and shape measurement noise will be taken into account later in forecasting the DESI-CSST performance. The galaxy clustering shotnoise measured in simulation is kept in variance, since it is an important ingredient of stochasticity. 

To make an intuitive comparison, we demonstrate $(\delta_A, \sigma_A)$ measurement for the full flux limited samples in cosmoDC2 galaxies in Fig.~\ref{fig:biasA_cosmoDC2-1} $\&$ \ref{fig:biasA_cosmoDC2-2}. We also present the results neglecting the impact of stochasticity, equivalently setting $r_s=1$, as the baseline of the systematic bias and the statistic uncertainty. In the context of amplitude $A$ quantification, the conclusions are same as that based on the ${\bf \bar{Q}}$. Some details are listed below.
\begin{itemize}
\item The statistic uncertainty decrease as redshift increase, as it is shown in the first row of Fig.~\ref{fig:biasA_cosmoDC2-1}, resulted from the larger cosmic volume therefore higher surface number density.
\item At redshift $z=1.0$ and $z=1.5$, the systematic bias of luminous populations is softened, counteracted by the enhanced sub-Poisson feature for the massive host halo population.
\item When an appropriate stochasticity parameterization is included in modeling, $\delta_A$ is alleviated while $\sigma_A$ is enlarged in different levels. The 2-parameter $r^2_s$ fits present a maximum benefit on constraining matter amplitude $A$, while the 1-parameter and 3-parameter fits do not achieve a competitive gain.
\item For 1-parameter fits, most of the galaxy samples retain bias at the level $\delta_A\gtrsim 2\%$ at $k_{\rm max}=0.9{\,\rm Mpc^{-1}}h$. The $\texttt{Quadratic-1}$ appears as the worst case, only few percent systematic bias is corrected, and there is still $\delta_A\sim 10\%$ at $z=0.5$. While for parameterizations such as $\texttt{Expan-gg-1}$ and $\texttt{Expan-mm-1}$, the systematic bias is controlled at the level $\delta_A\lesssim 5\%$ at $k_{\rm max}\simeq\, 1{\,\rm Mpc^{-1}}h$. The distinction is expected in the Eq.~(\ref{equ:stoch_pk}), which reveal that the stochasticity power spectrum is correlated with large scale clustering in higher order correlation terms, and the parameterizations such as $\texttt{Expan-gg-1}$ and $\texttt{Expan-mm-1}$ make use of the galaxy/matter clustering information while a naive low $k$ expansion $\texttt{Quadratic-1}$ makes flat assumption of stochasticity. 

\item For 2-parameter fits, the reconstruction accuracy is almost $\delta_A\lesssim 2\%$ at $k_{\rm max}=0.9{\,\rm Mpc^{-1}}h$ for $r^2$, and is generally $\delta_A< 2\%$ for $r^2_s$. However, for the luminous galaxy populations at $z=1.5$, a deviation $2\%\lesssim\delta_A\lesssim5\%$ appears in $r^2$ parameterizations including $P_{mm}$ (i.e. $\texttt{Expan-mm-2}$) but not in those including $P_{gg,s}$ (i.e. $\texttt{Expan-gg-2}$). We suppose the poor performance is caused by the strong halo exclusion in these massive host-halo populations, which contributes an enhanced scale dependent bias in non-linear region. The prominent non-linear bias is naturally included in the $P_{gg,s}$, but not related with $P_{mm}$. Fortunately, this issue does not affect the $r_s^2$ fits, because the overall halo exclusion always emerges as an upwarp in $\hat{r}_s^2$, which is well characterized by our $r^2_s$ parameterizations. 
\item For 3-parameter fits, $\delta_A$ is not corrected significantly compared to 2-parameter fits while $\sigma_A$ is magnified. For instance, when we replace parameterization $\texttt{Expan-mm-2}$ by $\texttt{Expan-mm-3}$ for the worst performing luminous galaxy populations, the correction of $\delta_A$ is improved by several percent, but $\sigma_A$ is amplified by about $20\%$.
\end{itemize}
The above discussion is based on cosmoDC2 samples, and the similar results are also obtained in the TNG300-1 galaxy samples.

In Table~\ref{table:SummarizeIndex}, we also show the mean value $\bar{\delta}_A$ and $\bar{\sigma}_A$ for the fiducial samples. Here the average value is defined by RMS, $\bar{\delta}_A = \sqrt{ \sum_i {\left(\delta_{A,i}\right)^2} }$ and $\bar{\sigma}_A = \sqrt{ \sum_i {\sigma_{A,i}^2} }$, where the summation index $i$ runs over all the galaxy samples. Now $\delta_A$ serves as a fair criterion to quantify the parameterization ability reconstructing the matter amplitude $A$ among different parameterizations. These summarized indexes reinforce above conclusions that the simple $\texttt{Quadratic}$ outperforms other parameterizations. Specifically, the recommended 2-parameter fits reach about $1\%\sim2\%$ accuracy of $\delta_A$, and approximate $3.5\%\sim 4\%$ precision of $\sigma_A$ for two sets of galaxy samples. Remarkably, $\texttt{Quadratic}$ scheme outperforms other schemes with a minimal $\delta_A=0.006$. What's more, we present the $\delta_A$-$\sigma_A$ relation scatter plot in Fig.~\ref{fig:summarizedIndex} for a visual comparison. 
\textcolor{black}{ We can find that the better reducing $\delta_A$ is accompanied by worse amplifying $\sigma_A$. For instance, for the parameterization schemes tested in the cosmoDC2 galaxies, the 3-parameter fit $\texttt{Expan-mm-3}$ alleviates the bias to $\bar{\delta}_A \sim 1\%$ with $\sim 5\%$ uncertainty. While the scheme $\texttt{Expan-mm-2}$ with less parameters has reduced performance with $\bar{\delta}_A\sim 2\%$ but shrinked uncertainty $\bar{\sigma}_A\sim 4.3\%$. One would obtain similar variation if further reducing 2 parameters to single one. Therefore, different schemes exhibit various behaviors, but the results suggest a negative correlation between the correction performance and the reconstruction uncertainty among similar parameterization schemes. }
To sum up, we exclude the 1-parameter $r^2$ fits on the right side in Fig.~\ref{fig:summarizedIndex} because of the large systematic bias, and exclude the 3-parameter $r^2$ fits and $\texttt{Q-bias}$ scheme on the top side because of the large uncertainty. The rest 2-parameter $r^2\,\&\,r^2_s$ schemes and $\texttt{Expan-mm-1}\;(r^2_s)$ scheme are applicable, in particular the $\texttt{Quadratic}$ scheme.

Apart from the intrinsic distinction in galaxy samples such as redshift, flux limit and mass, though not presented specifically, the simulation details also play an important role in stochasticity. Two galaxy catalogs utilized here exhibit obviously different behaviors. For instance, there is relatively subtle stochasticity for TNG300-1 samples compared to cosmoDC2, though with completely same selection criteria. Conclusions derived from two individual sets are consistent, but the performance details are different.


\begin{table}
    \renewcommand{\arraystretch}{1.3}
\caption{ \label{table:survey} Survey Parameters\footnote{
    Totally 7 redshift bins for cross correlation analysis. Here, $\Sigma_g$ /$n_g$ are the surface/volume number density of DESI LRG in each spec-z bin \cite{adame2023validation}, and $\Sigma_\gamma$ is the surface number density of CSST cosmic shear galaxy in each phot-z bin. }
 }
\begin{ruledtabular}
\begin{tabular}{ccc|cc}
    \multicolumn{3}{c|}{ DESI LRG } & \multicolumn{2}{c}{ CSST cosmic shear } \\ 
    $z$ & $\Sigma_g\,[{\rm deg^{-2}}]$ & $n_g\,[{\rm Mpc^{-3}h^3}]$ & $z_p$ & $\Sigma_\gamma\,[{\rm deg^{-2}}]$  \\ \hline
    $[0.4,\,0.5]$ & $47.5$ & $4.38\times 10^{-4}$ & $[0.7,\,3]$ & 68208 \\
    $[0.5,\,0.6]$ & $65.6$ & $4.49\times 10^{-4}$ & $[0.8,\,3]$ & 60462 \\
    $[0.6,\,0.7]$ & $80.0$ & $4.37\times 10^{-4}$ & $[0.9,\,3]$ & 52983 \\
    $[0.7,\,0.8]$ & $93.2$ & $4.25\times 10^{-4}$ & $[1.0,\,3]$ & 45959 \\
    $[0.8,\,0.9]$ & $99.3$ & $3.93\times 10^{-4}$ & $[1.2,\,3]$ & 33693 \\
    $[0.9,\,1.0]$ & $63.7$ & $2.24\times 10^{-4}$ & $[1.3,\,3]$ & 28528 \\
    $[1.0,\,1.1]$ & $28.3$ & $0.91\times 10^{-4}$ & $[1.3,\,3]$ & 28528 \\
\end{tabular}
\end{ruledtabular}
\end{table}

\begin{figure*}
\includegraphics[width=1.0\textwidth]{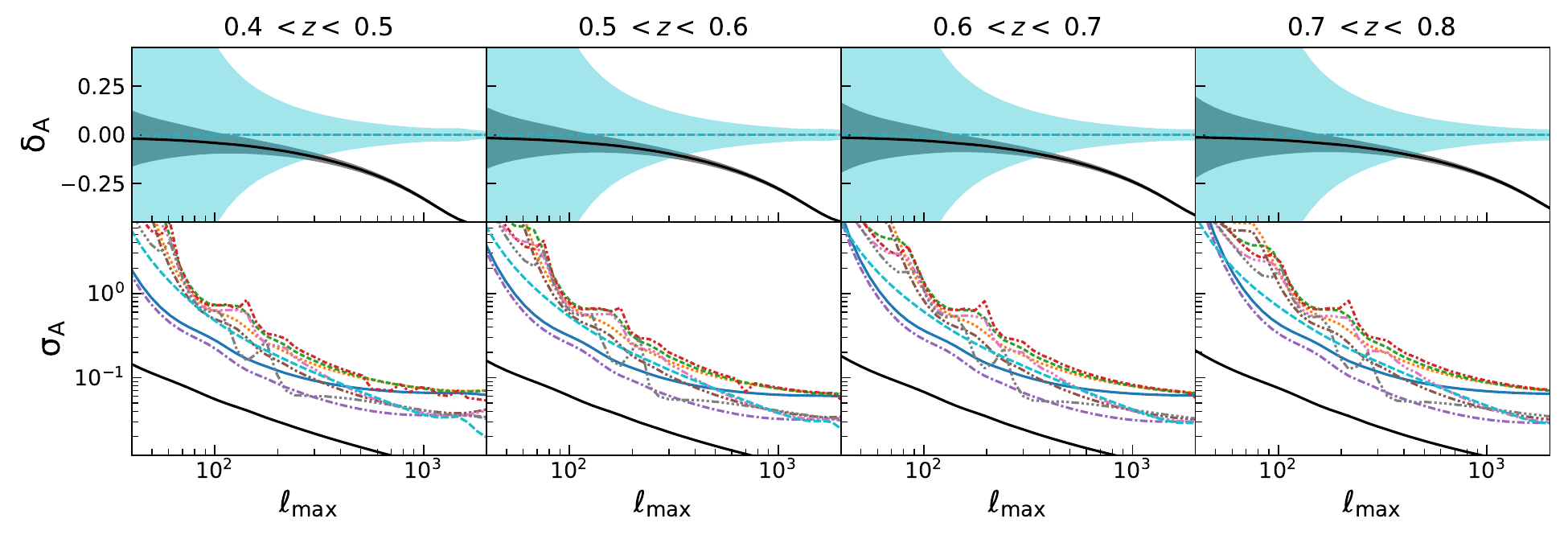}
\includegraphics[width=1.0\textwidth]{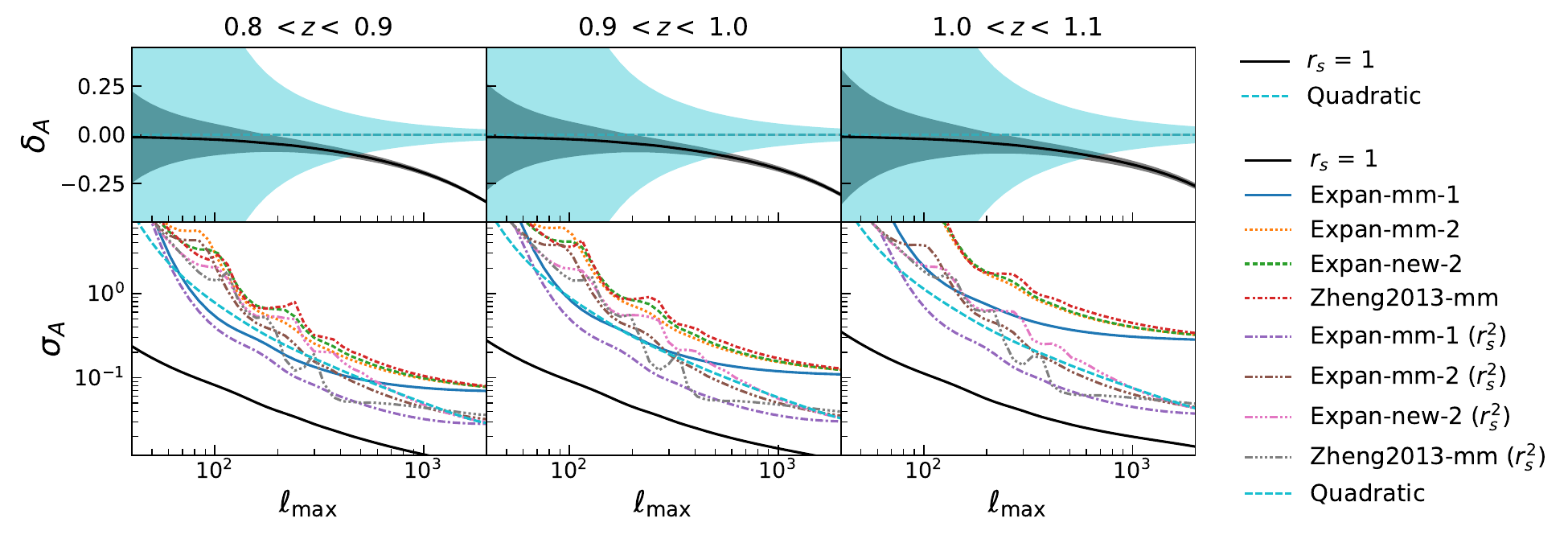}
\caption{ \label{fig:forecast_desi} Forecasted stochasticity impact on the matter amplitude uncertainty $\sigma_A$ for DESI-like galaxy $\times$ CSST-like cosmic shear. In both top and bottom panels, 7 columns show the results at different redshift bins. The top subfigures show systematic bias $\delta_A$ by assuming $r_s=1$ (black line) and the fiducial case $A=1$ with $r_s^2$ parameterized by $\texttt{Quadratic}$ (cyan line), and the $1\sigma$ regions are marked. The bottom subfigures show statistic uncertainty $\sigma_A$ under various parameterizations compared to the case $r_s=1$ (black line). The cosmological parameters are fixed as dark matter simulation described in Appendix~\ref{sec:appendix_samples}. }
\end{figure*}

\begin{figure}
\includegraphics[width=1.0\columnwidth]{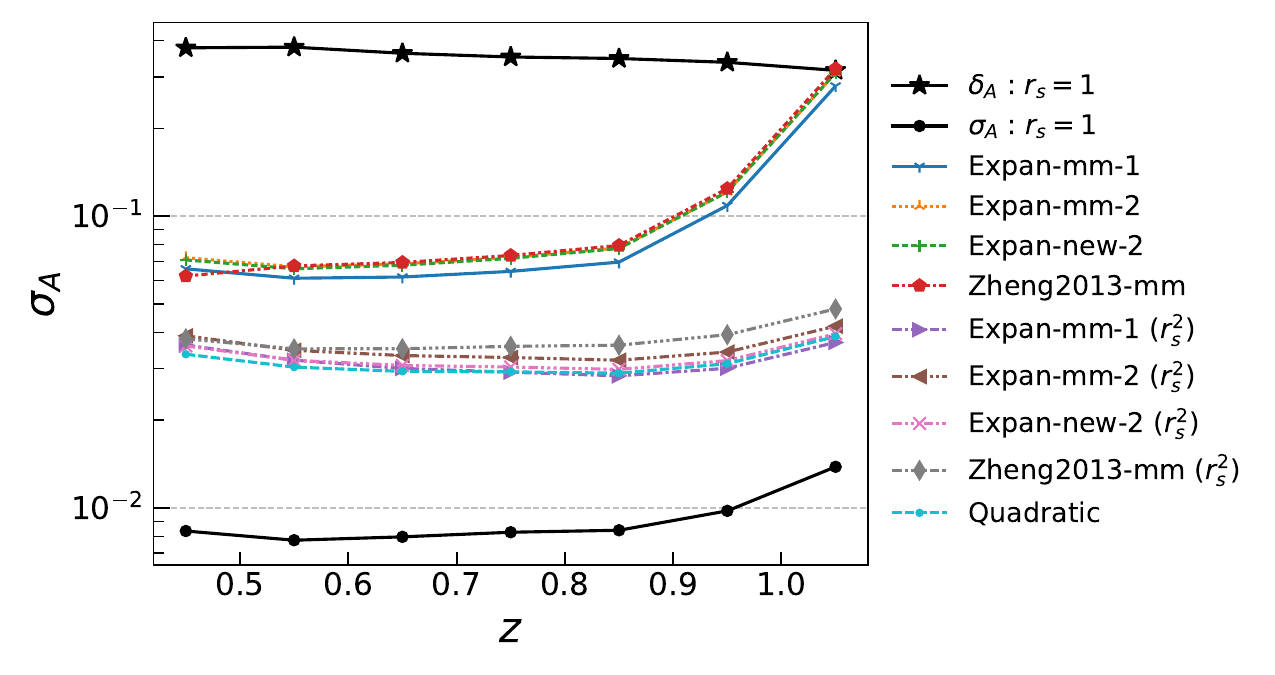}
\caption{ \label{fig:forecast_desi_asz} Forecasted statistic uncertainty $\sigma_A$ as function of redshift bin $z$ for DESI-like galaxy $\times$ CSST-like cosmic shear. The result shown here is fixed with $\ell_{\rm max}\chi_g=0.9 \,h^{-1}{\rm Mpc}$.  }
\end{figure}

\begin{figure}
\includegraphics[width=1.0\columnwidth]{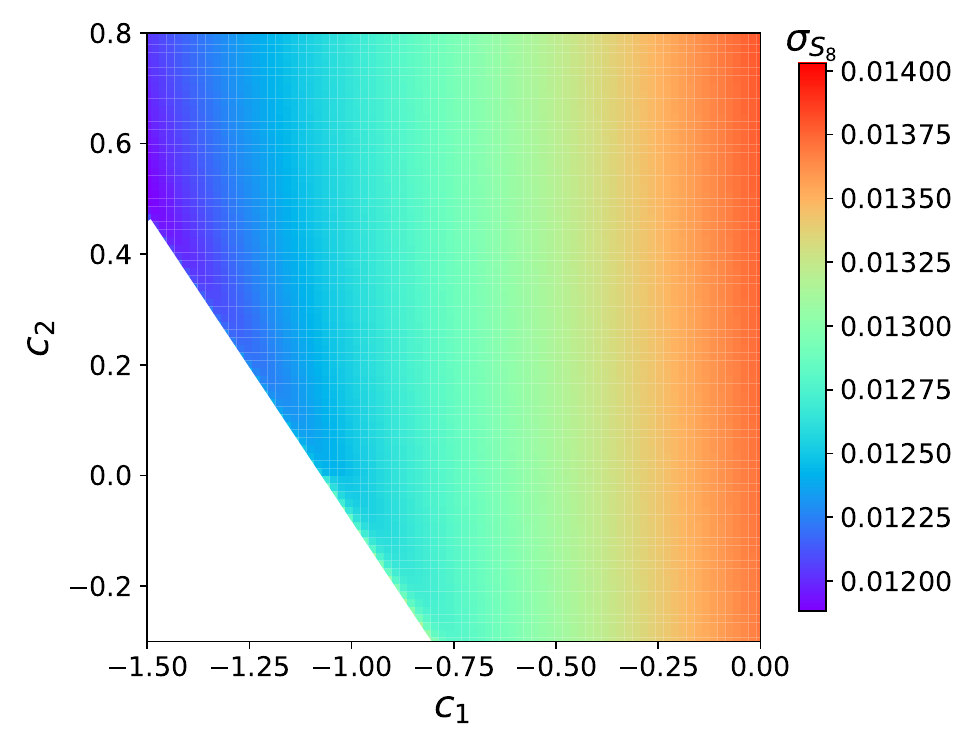}
\caption{ \label{fig:S8_sigma2} Forecasted statistic uncertainty of cosmological parameter $S_8$ as function of stochasticity parameters $(c_1, c_2)$. The blank region in left bottom is null value because of unphysical $r^2<0$ for these parameters.  }
\end{figure}

\section{Forecast for DESI-like LRG \label{sec:desi-like}}

The stochasticity parameterization alleviates the systematic bias to a large extent, but including nuisance parameters reduces the power of cosmological parameters constraint. Here we forecast the specific impact on the combination of DESI-like galaxy and CSST-like cosmic shear. We adopt $\texttt{Quadratic}$ scheme as the fiducial stochasticity behavior for DESI-like galaxies, because it has been shown to present the best performance among all candidates.

The forecasted surface number density of DESI-like galaxies and the estimation of CSST-like cosmic shear survey are summarized in Table.~\ref{table:survey}. For the DESI spectroscopic tracer, we only utilize LRG in our analysis. Though the previous galaxy samples selection differs from DESI LRG selecting strategy \cite{Zhou_2023}, it is supposed to affect little on our forecast. On the one hand, DESI LRG sample is selected using both magnitude and color, approximately being flux limited samples for which we have performed various tests. On the other hand, our parameterizations have been shown to be powerful enough to characterize various kinds of stochasticity, so it is expected to be applicable to realistic DESI LRG. The validation and the forecast for the DESI ELG samples, the major targets with complex identification \cite{Raichoor_2023}, is remained to future investigation. In this work, LRG samples are divided into $7$ redshift bins covering redshift from $0.4$ to $1.1$ with $\Delta z=0.1$. 
{\color{black} We adopt the linear galaxy bias $b(k,z)\equiv b_L(z)=1.7/D(z)$, where $D(z)$ is the linear growth factor \cite{adame2023validation}. }

For the cosmic shear tracers, CSST photometric galaxy survey is expected to reach a total galaxy surface number density $\Sigma_g=28\,{\rm arcmin^{-2}}$ with redshift scatter $\sigma_z=0.05$. In our analysis, the redshift distribution is adopted as $n(z)\propto z^2e^{-z/z_*}$ covering redshift $0<z<4$, where $z_*=0.35$. The photometric redshift bins are selected to avoid the overlapping of spectroscopic galaxy and the cosmic shear tracers, as well as minimize the shotnoise of $\kappa$ field. We assume the Gaussian variance for CSST-like cosmic shear, and adopt the estimated  shape noise $\sigma_\gamma=0.2$ and additive noise $N_{\rm add}^\gamma=10^{-9}$ \cite{gong2019cosmology}. The overlapping fraction of sky with DESI-like galaxies is $f_{\rm sky}=0.3$. To avoid complexity, we have neglected the impacts of magnification bias, photometric error and other systematic uncertainties.

For illustrative purpose, we simply make an aggressive guess of stochasticity with $\texttt{Quadratic}$ parameters $c_1=-1.57$, $c_2=0.74$ for all 7 galaxy bins (called Scenario-A below), and treat all $2\times 7$ nuisance stochasticity parameters as free parameters when applying fisher forecast.
We do not show the results for a scheme such as $\texttt{Expan-gg-2}$, because it is equivalent with $\texttt{Expan-mm-2}$ under the linear bias assumption. 
We show the results under Scenario-A stochasticity in Fig.~\ref{fig:forecast_desi}. If we neglect the impact of stochasticity and adopt $r_s=1$, the systematic bias $\delta_A$ would exceed the statistic uncertainty at $k_{\rm max}\equiv\ell_{\rm max}\,\chi_g\simeq 0.1{\,\rm Mpc^{-1}}h$, where $\delta_A\simeq\sigma_A\simeq0.05$. As $k_{\rm max}$ extends to $0.9{\,\rm Mpc^{-1}}h$, as shown in Fig.~\ref{fig:forecast_desi_asz}, the uncertainty of $r_s=1$ is reduced to $\sigma_A\simeq 0.01$ while the underestimation is serious $\delta_A=0.3\sim0.4$. When we adopt parameterization $\texttt{Quadratic}$, the matter clustering amplitude is estimated unbiasedly with precision $\sigma_A \simeq 0.04$ for the first 5 bins and $\sigma_A \simeq 0.045,\;0.059$ for the latter 2 bins.
In the bottom subfigures of each column in Fig.~\ref{fig:forecast_desi}, we further compare the abilities of other parameterization schemes. 
Among these various schemes, the values of $\sigma_A$ bifurcate into two branches as $\ell_{\rm max}$ increase. One branch has larger $\sigma_A$ for $r^2$ parameterizations, while another branch has smaller $\sigma_A$ for $r^2_s$ parameterizations. The reason for these distinctive behaviors of two kinds of parameterization schemes is the flattened effect of the discrete noise in galaxy power spectrum. If we do not subtract the Poisson noise expectation when $\bar{n}_gP_{gg}\ll 1$, the curve of $r^2$ is suppressed to a nearly flat shape, that $r_s>r$. Thus the modeling $\hat{A}=rA$ is less sensitive compared to the modeling $\hat{A}_s=r_s A$, quantitatively,
$ |\frac{\partial \hat{A}_s}{\partial \lambda}|> |\frac{\partial \hat{A}}{\partial \lambda}|$,
where $\lambda$ is the cosmological parameter we interest in. It leads to a larger uncertainty of $A$ in $r^2$ modeling. Consistent with the argument, the bifurcation is significant for the last redshift bin, of which the number density is lowest, reaching $\bar{n}_gP_{gg,s}=0.1$ at $\ell_{\rm max}\simeq 1500$. For the relative higher number density bin, where $\bar{n}_gP_{gg,s}=0.1$ at $\ell_{\rm max}\simeq 3200\sim 5000$, the bifurcation is mild. 

The underestimation in matter clustering amplitude would propagate to cosmological constraint and result into a substantial effect, sourcing a potential systematic bias in cosmological parameters.
Different experiments have reported the tendency that the lower redshift measurements prefer lower matter clustering amplitude compared with the CMB constraint in the context of $\Lambda{\rm CDM}$ \cite{krolewski2021cosmological, white2022cosmological, abbott2022dark, marques2024cosmological}. For the cosmic shear measurement, this anomaly is usually summarized as the structure growth parameter $S_8\equiv\sigma_8(\Omega_m/0.3)^{0.5}$. Since the value of $S_8$ (or $\sigma_8$) is degenerated with $r$, an appropriate $r^2$ modeling excludes the potential systematic bias thus obtain an unbiased constraint.
We demonstrate the stochasticity impact on cosmological constraint by varying the parameters $\Omega_m$ and $\sigma_8$, and present the constraint performance for $S_8$ together with/without stochasticity parameterization. 
We find:
\begin{itemize}
\item In the Scenario-A, an aggressive guess of serious stochasticity, our fiducial results with $k_{\rm max}=0.9{\,\rm Mpc^{-1}}h$ are $\sigma_8=0.83\pm0.03$ and $\Omega_m=0.268\pm0.012$, with precision $3.5\%$ and $4.5\%$ respectively. The structure growth is $S_8=0.784\pm0.012$, with precision $1.5\%$. 
\item In the Scenario-A, the neglect of stochasticity imports significant bias on cosmological parameter. Setting $r_s=1$ and even at the linear scale $k_{\rm max}=0.1{\,\rm Mpc^{-1}}h$, the inferred $S_8=0.746\pm0.012$ deviates from the fiducial values $S_8$ on the level of $3\sigma$. The deviation is moderate if we directly constrain $\sigma_8$, which is only about $0.3\sigma$ deviated from fiducial value at $0.1{\,\rm Mpc^{-1}}h \leq k_{\rm max} \leq 0.3{\,\rm Mpc^{-1}}h$. But the $\sigma_8$ precision is limited to $>3\%$ at $k_{\rm max}<0.3{\,\rm Mpc^{-1}}h$ due to the degeneration of $\sigma_8$ and $\Omega_m$, and its systematic bias reaches about $3\sigma$ again at $k_{\rm max}\gtrsim 0.4{\,\rm Mpc^{-1}}h$.
\item In the Scenario-B, a mild stochasticity hypothesis with $c_1=-0.63, c_2=0.27$, the fiducial results are $\sigma_8=0.83\pm0.03$, $\Omega_m=0.268\pm0.012$ and $S_8=0.784\pm0.013$, with constraint precision $3.4\%,\,4.5\%$ and $1.7\%$ respectively. The precision varies slightly compared with the Scenario-A, though we adopt quite different $r_s^2$ value. 
\item In the Scenario-B, the deviation is mitigated to a large extent compared to the Scenario-A, but when we set $r_s=1$, the true value of $S_8$ is still outside the $2\sigma$ region of inferred values at $k_{\rm max}=0.3{\,\rm Mpc^{-1}}h$. As $k_{\rm max}$ increases, the deviation is more and more serious, obtaining $S_8=0.736\pm0.007$ at $k_{\rm max}=0.5{\,\rm Mpc^{-1}}h$, a $7\sigma$ deviation. While for the $\sigma_8$ constraint, the systematic bias is not significant compared with statistic fluctuation, but we still have $\sigma_8=0.79\pm0.02$ at $k_{\rm max}=0.5{\,\rm Mpc^{-1}}h$, a $2\sigma$ deviation. 
\item In Fig.~\ref{fig:S8_sigma2}, we present the statistic uncertainty for the fiducial forecasted $S_8$ by varying the stochasticity parameters. In the parameter space, the variation of $(c_1, c_2)$ introduces distinguished behavior of stochasticity, as well as suppression of matter clustering amplitude in different levels. But the statistic uncertainty for the final constrained $S_8$ only varies in a narrow range $0.012 \lesssim\sigma_{S_8}\lesssim 0.014$. We suppose the uncertainty is limited by the large number of constraint parameters (total 14 nuisance parameters plus 2 cosmological parameters). Thus there are tight bounds for $\sigma_{S_8}$ with various $(c_1, c_2)$, rather than a strong dependence on the specific shape of $r_s^2$. Similar tight bounds arise for both $\Omega_m$ and $\sigma_8$ uncertainty. Consequently, the uncertainty of the inferred parameters is controlled in a limited range, not matter what kinds of stochasticity.
\end{itemize}

We aim to provide an unbiased cosmological parameters prediction, but one may wonder that an accurate stochasticity model seems to pay a high price in constraining power. We suppose this is not an issue for the following reasons. Firstly, the vast parameter set (14 nuisance parameters for 7 redshift bins) is not necessary in realistic case, since $\Delta z$ is so small that we expect the clustering as well as stochasticity slowly evolves relative to adjacent redshift bins. Thus some simple continuity hypothesis is enabled and the number of nuisance parameters is reduced. Secondly, with the assistance of calibration in the targeted mock, some proper priors of stochasticity parameters are available \cite{kokron2022priors}, then reducing the uncertainty. Finally, the stochasticity components are considerable compared to clustering signal in current spectroscopic survey, such as Ref.~\cite{bose2023euclid, pezzotta2023euclid}. And stochasticity modeling is also important to obtain a consistent result in the future multi-probes applicants \cite{alexander2020testing,pezzotta2021testing}. Thus the inclusion of stochasticity nuisance parameters is inevitable, and the key, also our future work, is how to further reduce the degree of freedom in stochasticity parameters for tomographic analysis.


\section{Conclusion and Discussions}

In this work, we addressed two leading questions: how to parameterize stochasticity in galaxy clustering, and how to reconstruct the unbiased tomographic matter clustering.
In order to seek an appropriate phenomenological description of stochasticity, we investigate 12 kinds of parameterization schemes for the cross correlation coefficients between galaxy overdensity $\delta_g({\bf k})$ and the underlying matter overdensity $\delta_m({\bf k})$.
We test the performance against the galaxy samples selected from cosmoDC2 and TNG300-1 simulations, over a wide range of redshift, flux and various galaxy colors. We quantify the fitting performance by the absolute difference of $r_{(s)}(k)$ values and the accuracy $\&$ precision of reconstructed matter clustering amplitude. Then we choose the best performing quadratic scheme as the fiducial stochasticity, and present the fisher forecast of the matter clustering amplitude reconstruction as well as the cosmological parameters constraint, assuming the combination DESI-like LRG and CSST-like cosmic shear. Our main conclusions are as following:

\begin{enumerate}[(i)]
\item We verify the suggestion the effective ${\rm DoF}\simeq 2$ in galaxy stochasticity, and that 2-parameter scheme provides the maximum gain in the stochasticity description compared to 1-parameter or 3-parameter scheme. Against two sets of galaxy samples, we have shown that 2-parameter fits are able to reach about $1\%$ accuracy for $r_{(s)}^2$, and obtain general $0.6\%\sim2\%$ accuracy for the reconstructed matter clustering amplitude at $k_{\rm max}=0.9{\,\rm Mpc^{-1}}h$.
\item The quadratic scheme $r^2_s = 1+c_1 k+c_2 k^2$ is the most promising parameterization among all the schemes, performing the best direct fit for $\hat{r}_s^2$ and reconstructing the matter clustering amplitude with better than $1\%$ accuracy. 
\item For the combination DESI-like LRG $\times$ CSST-like cosmic shear, we forecast that there is a suppression in the reconstructed tomographic matter clustering amplitude when neglecting the stochasticity. The systematic bias sourced by the suppression is comparable with statistic uncertainty at linear scale $k_{\rm max}=0.1{\,\rm Mpc^{-1}}h$ if the stochasticity contamination is serious. The serious stochasticity also biases the inferred $S_8$ at $3\sigma$ level even at $k_{\rm max}=0.1{\,\rm Mpc^{-1}}h$. 
\item Together with the quadratic stochasticity parameterization, the combination DESI-like LRG $\times$ CSST cosmic shear is expected to achieve $S_8$ constraint at $1.5\%$ precision, free from the stochasticity systematic. The uncertainty is likely to be further reduced with stochasticity parameter number reduced.
\end{enumerate}
We caution that this work is still theoretical. Its applicability in real data requires validation in targeted mocks, which must be done case to case. 

The galaxy-matter cross power spectrum is also available by the cross correlation with CMB lensing. The CMB lensing is much cleaner than the cosmic shear estimated from the galaxy ellipticity, and the photometric redshift distribution of lensed galaxies introduces large uncertainty while the CMB redshift is accurately known. Thus we expect the combination with galaxy clustering $\times$ CMB lensing achieves a tighter tomographic matter clustering constraint \cite{peacock2018wide}. 
Apart from the direct reconstruction of tomographic matter clustering from galaxy/CMB lensing, there are wide applications for an accurate $r^2$ parameterization. For example, it could alleviate the systematic bias due to stochastic components in the estimator of $D_G$ statistic \cite{giannantonio2016cmb, omori2019dark, marques2020tomographic, marques2024cosmological}. In the condition of non-negligible stochasticity, the theoretical expectation of estimator $\hat{D}_G$ (refer to Ref.~\cite{giannantonio2016cmb} for detailed expression) can be rewritten to correct the stochasticity suppression, $\langle \hat{D}_G(\ell)/ r\left(\frac{\ell}{\chi}\right) \rangle = D(z)$, where $D(z)$ is the linear growth function.

We plan to validate our methods with the simulations designed for DESI spectroscopic redshift surveys, in particular the ELG mocks. After the validation, we will apply our methods to the DESI spectroscopic sources to reconstruct the matter power spectrum. Meanwhile, our analysis simultaneously provides the stochasticity parameters for the galaxy samples, therefore we can determine whether the stochasticity is a serious systematic bias in the joint analysis of galaxy auto- and cross-power spectrum or not. It would serve as an important supplementation for systematic diagnosis in the future DESI multi-probes applications. We expect our reconstructions will be complementary to DESI constraint of dark energy and modified gravity by BAO and RSD measurement.

\begin{acknowledgments}
We would like to thank Yu Yu's help in the access and usage of simulations, and thank Hong-Ming Zhu for kind suggestions. This work is supported by  the National Key R\&D Program of China (2023YFA1607800,2023YFA1607801,2020YFC2201602), the National Science Foundation of China (11621303), CMS-CSST-2021-A02, and the Fundamental Research Funds for the Central Universities. This work made use of the Gravity Supercomputer at the Department of Astronomy, Shanghai Jiao Tong University. 

We acknowledge the use of code package $\mathtt{nbodykit}$ \cite{Hand_2018}, $\texttt{CAMB}$ \cite{2011ascl.soft02026L} and $\texttt{CCL}$ \cite{Chisari_2019}.

\end{acknowledgments}


\appendix


\begin{figure*}
\includegraphics[width=1.0\textwidth]{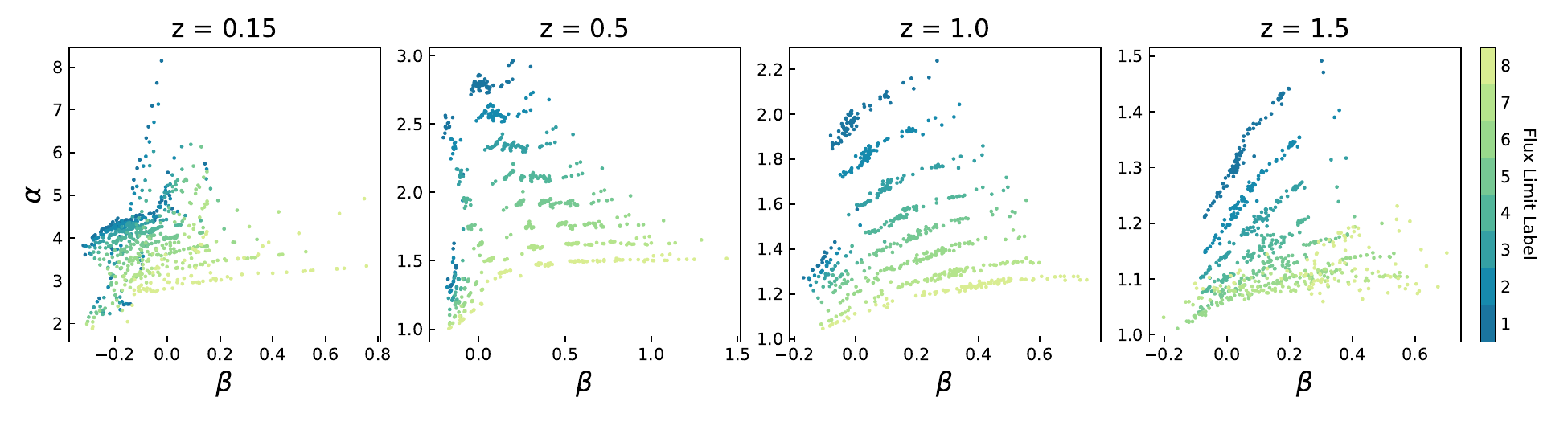}
\includegraphics[width=1.0\textwidth]{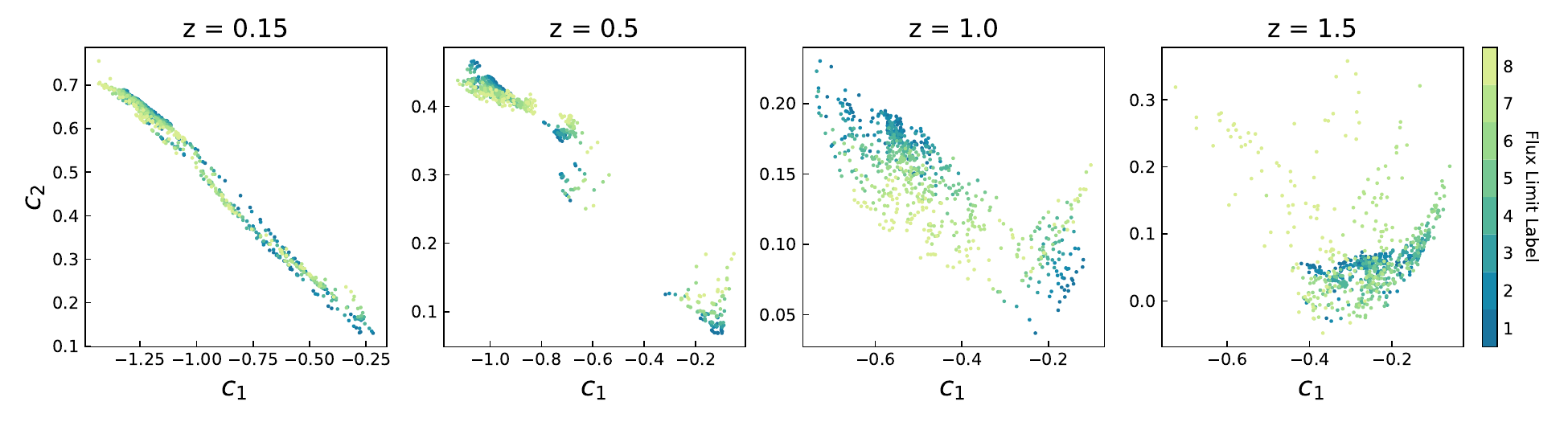}
\caption{ \label{fig:add_1} Scatter plots of parameters distribution in 4 redshift snapshots based on cosmoDC2 samples. The top panels show parameters $(\alpha,\beta)$ in $\texttt{Expan-mm-2}$, and the bottom panels show parameters $(c_1, c_2)$ in $\texttt{Quadratic}$, with $k_{\rm max}=0.9{\,\rm Mpc^{-1}}h$. In each subfigure, all various galaxy color-bins with same $\mathsf{FluxLimitLabel}$ are marked with same color.  }
\end{figure*}

\begin{figure}
\includegraphics[width=1.0\columnwidth]{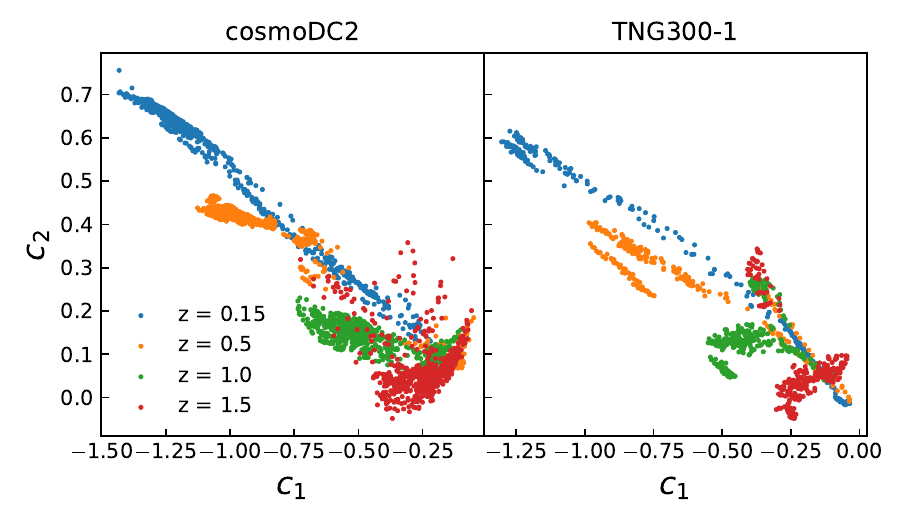}
\caption{ \label{fig:add_2} Comparison of parameters $(c_1, c_2)$ in $\texttt{Quadratic}$ between cosmoDC2 and TNG300-1 galaxy samples. In each subfigure, all different $\mathsf{FluxLimitLabel}$ and different galaxy color-bins in same redshift are marked with same color, with $k_{\rm max}=0.9{\,\rm Mpc^{-1}}h$.  }
\end{figure}


\section{Galaxy Samples \label{sec:appendix_samples}}

The performance testing for the parameterizations in Table~\ref{table:formula} is based on the fiducial color-bins samples, where the specific definitions of various galaxies refer to Table.~\ref{table:galaxy_cosmoDC2} and Table.~\ref{table:galaxy_TNG300}. The fiducial galaxy samples consist of two sets of galaxy samples, cosmoDC2 mock \cite{korytov2019cosmodc2} and TNG300-1 galaxy samples from the IllustrisTNG project \cite{Springel_2017, Nelson_2017, Pillepich_2017, Naiman_2018, Marinacci_2018}. The cosmoDC2 is a synthetic galaxy catalog generated for LSST survey \cite{korytov2019cosmodc2}, and it provides a lightcone covering $440\,{\rm deg}^2$ of sky area, with $\sim2.6$ billions of galaxies distributed to redshift $z=3$. The magnitude $ugrizY$ photometirc bands and the host halo are provided for the galaxies. 
The hydrodynamic simulation TNG300-1 is a publicly available simulation from the IllustrisTNG project \cite{Springel_2017, Nelson_2017, Pillepich_2017, Naiman_2018, Marinacci_2018}. IllustrisTNG is a suite of large volume, cosmological, gravo-magnetohydrodynamical simulations run with the moving-mesh code ${\tt APEPO}$ \cite{springel2010pur}. IllustrisTNG implements a set of physical processes to model the galaxy formation process including gas cooling, star formation and evolution, supernova feedback \cite{pillepich2018simulating}, and AGN feedback \cite{weinberger2016simulating}. This project includes TNG50, TNG100 and TNG300, with boxsize of $50$ Mpc, $100$ Mpc and $300$ Mpc respectively. We choose TNG300-1 for our analysis for its largest amount of galaxy samples. We identify all luminous subhalos in the TNG300-1 as galaxies.

However, the accessed cosmoDC2 products do not included the associated 3D matter field, which is based on the Outer Rim simulation \cite{Heitmann_2019}. Regarding the TNG300-1 simulation, the products we assess are complete but the boxisize $L\,=\,205\,{\rm Mpc/h}$ is limited for our analysis, which lacks the statistic for linear scale $k\lesssim 0.1 {\rm\,Mpc^{-1}}h$ and induces large fluctuation in $r^2$. To solve these problems, we combine the galaxies in cosmoDC2$\;\&\;$TNG300-1 with dark matter CosmicGrowth simulation \cite{jing2019cosmicgrowth}. The CosmicGrowth simulation adopts a flat WMAP cosmology \cite{hinshaw2013nine}, of which the boxsize is $L\,=\,1200\,{\rm Mpc/h}$ and particle number is $3072^3$. The cosmological parameters adopt $\Omega_b=0.0445\,,\;\Omega_c=0.2235\,,\;h=0.71$ and $\sigma_8=0.83$. We use the galaxy-host halo relation of the galaxy in cosmoDC2$\;\&\;$TNG300-1 to assign galaxies to the CosmicGrowth simulation halos, preserving the conditional probability of galaxy contents given hosthalo mass \cite{korytov2019cosmodc2}. Because of the distinction of these simulations resulted from different cosmological parameters and the simulation details, we do an abundance matching between cosmoDC2$\;\&\;$TNG300-1 and CosmicGrowth halos, and then assign galaxies in cosmoDC2$\;\&\;$TNG300-1 to CosmicGrowth halos.


\begin{figure}
\includegraphics[width=1.0\columnwidth]{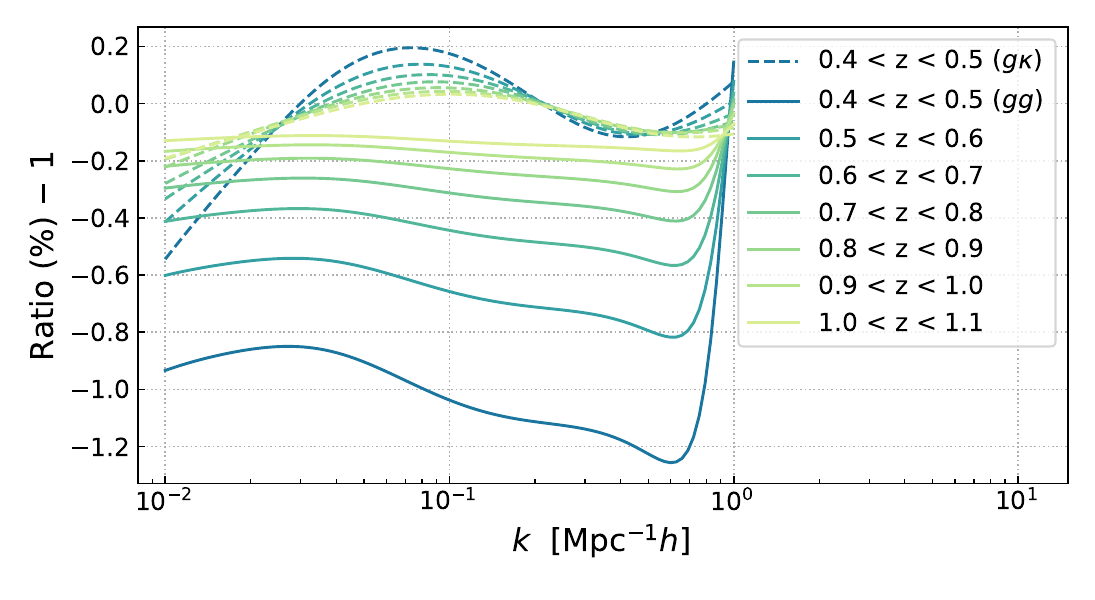}
\includegraphics[width=1.0\columnwidth]{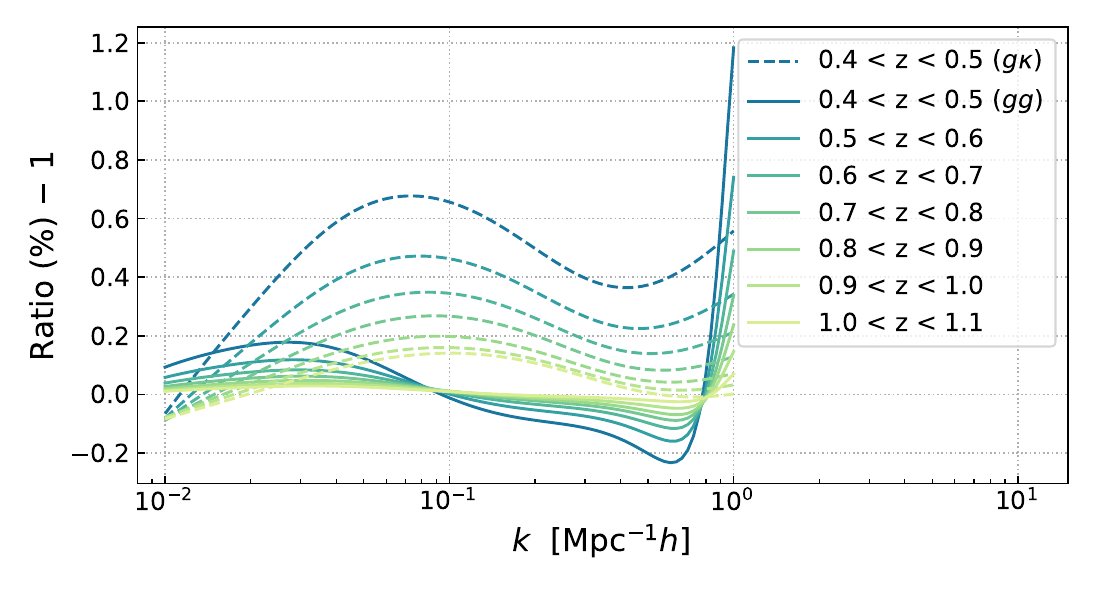}
\caption{ \label{fig:append_formulaAccuracy}  \color{black}
The ratio difference of the 3D clustering at effective redshift to the projected integral of full redshift distribution. 
In the upper figure, we present the ratio difference of Eq.~(\ref{equ:Cgkappa_simplied}) and Eq.~(\ref{equ:Cgg_simplied}), specifically,  $\chi_g^{-2}\,W_L\, \hat{P}_{gm} \,/\, \hat{C}^{g\kappa}_\ell $ in dash line and $\chi_g^{-2}\,\Delta\chi_g^{-1}\, \hat{P}_{gg} \,/\, \hat{C}^{gg}_\ell$ in solid line. 
In the lower figure, we present the ratio difference of Eq.~(\ref{equ:C_gkappa_approx1}) and Eq.~(\ref{equ:C_gkappa_approx2}), specifically,  $\left[ \int{ d\chi \,\chi^{-2}\, n_g(\chi) W_L(\chi) }\right] \hat{P}_{gm}\,/\, \hat{C}^{g\kappa}_\ell $ in dash line and $\left[\int{ d\chi \,\chi^{-2}\, n_g(\chi)n_g(\chi) }\right] \hat{P}_{gg} \,/\, \hat{C}^{gg}_\ell$ in solid line. 
Here, we assume the volume number density of the galaxy samples is constant within each redshift bin, which is expected for DESI LRG samples at $0.4 < $z$ <0.8$ \cite{Zhou_2023, adame2024desi}. 
The galaxy bias $b(z)$ and stochasticity (Scenario A) adopted in Section~\ref{sec:desi-like} are considered appropriately in both $\hat{P}_{gm}$ and $\hat{P}_{gg}$ calculation. 
}
\end{figure}

{ \color{black}
\section{Redshift Tomography Approximation}
\label{append:ApproxiAccruracy}

In the main text, we utilize the approximations of Eq.~(\ref{equ:Cgkappa_simplied}) and Eq.~(\ref{equ:Cgg_simplied}) in order to present a clean and intuitive derivation for the principles of performance test. To reach out to the simplicity, we have assumed the galaxy redshift distribution is narrow enough that we can approximate $n(\chi)$ as $\delta^D(\chi-\chi_g)$. Furthermore, we have also assumed its asymptotic behavior from normalization condition,
\begin{equation}
\begin{split}
\int n_g(\chi) d\chi=1  &\quad\to\quad   n_g(\chi)\Delta\chi_g = 1  \\
{\rm\qquad for\quad} & \chi_g-{\Delta\chi_g\over 2} < \chi < \chi_g+{\Delta\chi_g\over 2}   \quad,
\end{split}
\end{equation}
which is exactly true if the galaxy redshift distribution is proportional to $H^{-1}(z)$, where $H(z)$ is the Hubble parameter. The effective redshift $\hat{z}_g$ is adopted as the mean value of the galaxy distribution \cite{bautista2021completed}, $\hat{z}_g = \int z\, n_g(z)\,dz$. Rigorously, the approximation introduces potentially sub-percent level of inaccuracy when we adopt $\Delta z=0.1$, and we should validate the possible impact in realistic applications. However, it does not impact on the estimation results in the main text, since the shear noise dominates the uncertainty. 

In the upper figure of Fig.~\ref{fig:append_formulaAccuracy}, we present the estimation of the ratio difference for 7 spec-z bins listed in Table~\ref{table:survey}. There is only sub-percent level of difference $\lesssim 1\%$ for all 7 bins, and the ratio difference decreases significantly for higher redshift bins due to the reduced evolution effect compared to lower redshift bins. In principle, we can adopt narrower redshift bins to further reduce this minor effect, for instance, a redshift bin with width $\Delta z = 0.05$ can reduce the ratio difference to $\lesssim 0.3\%$. Additionally, we can further mitigate the mismatch of galaxy redshift distribution and lensing kernel by weighting the galaxy with $w(z) = W_L(\chi)/n_g(\chi)$ in both auto- and cross-power spectrum \cite{chen2022cosmological}.

A more conservative method assumes the matter and galaxy clustering evolve slowly compared the variation of integral kernel within the bin, and then we can approximate Eq.~(\ref{equ:C_gkappa_measured}) as
\begin{eqnarray}
\hat{C}^{g\kappa}_\ell &=& \left[ \int{ d\chi \,\chi^{-2}\, n_g(\chi) W_L(\chi) } \right] \hat{P}_{gm}\left( k, \hat{z}_g \right)    \label{equ:C_gkappa_approx1}  \\
\hat{C}^{gg}_\ell &=& \left[ \int{ d\chi \,\chi^{-2}\, n_g(\chi)n_g(\chi) }  \right]  \hat{P}_{gg}\left( k, \hat{z}_g \right)    \label{equ:C_gkappa_approx2}  
\end{eqnarray}
which account for the slight variation of the integral kernel and only modify the prefactor compared to the approximation Eq.~(\ref{equ:Cgkappa_simplied}) and Eq.~(\ref{equ:Cgg_simplied}). For $k<0.9\,{\rm Mpc}^{-1}h$, it also has $< 1\%$ ratio difference with $\Delta z=0.1$, as shown in the lower figure of Fig.~\ref{fig:append_formulaAccuracy}. What's more, all the results shown in Fig.~\ref{fig:append_formulaAccuracy} are not sensitive to the assumption of galaxy bias and stochasticity.

While for photometric objects, the accurate redshift tomography combining galaxy and lensing is a highly challenging task due to the intrinsic redshift uncertainty, finite redshift width, and the mismatch between galaxy distribution and lensing kernel \cite{abbott2022dark, zhou2024challenges}. Thus, rather than directly reconstructing the 3D clustering, forward modeling is a more preferred solution \cite{krolewski2020unwise,white2022cosmological,karim2024measuring}. 

}


\section{Gaussian Variance}

The direct fitting of the measured $r^2$ in the simulations to our parameterized formula is through minimizing the Eq.~(\ref{equ:chi2_fitting}), where the covariance is assumed to be Gaussian. Start with the covariance of power spectrum of Gaussian fields $i,\,j,\,p,\,q$, 
\begin{eqnarray}
    {\rm\bf Cov}\left[P_{ij}(k), P_{pq}(k')\right] = \frac{\delta_{k, k'}^{K}}{N_{\bf k}}\left( P_{ip}P_{jq} + P_{iq}P_{jp} \right) \; ,
\end{eqnarray}
it is straightforward to derive the Gaussian variance of $r^2$ and $r^2_s$, 
\begin{eqnarray}
    \label{equ:gaussion_variance_r2}
    \sigma^2|_{r^2} &=& \frac{4}{N_k} \left(r^2\right)^2 \left( \frac{1}{r^2} + r^2 -2 \right) \quad , \\
    \sigma^2|_{r^2_s} &=& \frac{4}{N_k} \left(r_s^2\right)^2 \left[ \frac{1}{r_s^2} + r_s^2 -2  \right. \nonumber\\
    & & + \left. \frac{1}{\bar{n}_g P_{gg,s}} \left(\frac{1}{r_s^2}-1\right) +\frac{1}{2\left(\bar{n}_g P_{gg,s}\right)^2} \right]  \quad . \nonumber
\end{eqnarray}


\section{Parameter Distribution}

In Fig.~\ref{fig:add_1}, we demonstrate the parameters distribution for two schemes $\texttt{Expan-mm-2}$ and $\texttt{Quadratic}$. As shown in the figure, stochasticity parameters exhibit certain relevance with respect to the galaxy properties, and this is a natural outcome. On the one hand, the stochasticity traces the complex nature of galaxy distribution, thus they should exhibit regular evolution to varying degree in parametric space. On the other hand, as it is indicated by Eq.~(\ref{equ:stoch_pk}), the stochasticity is coupled with large scale structure fluctuation in higher order correlation. The coevolution effect also promotes the stochasticity parameters to behave regularly. For the $(\alpha,\beta)$ parameters in $\texttt{Expan-mm-2}$, the bins in different $\mathsf{FluxLimitLabel}$, equivalently different $\bar{n}_g$, are separated into different tracks. It can be understood by the form of $\texttt{Expan-mm-2}$, where the parameter $\alpha$ is degenerated with $\bar{n}_g$ (as well as galaxy bias in fact), thus any variation in stochasticity corresponding to number count will propagate to $\alpha$ directly though $\bar{n}_g$. It turns out as that different galaxy samples populations are offset according to $\mathsf{FluxLimitLabel}$, modulate by $\alpha$. Such obvious separation does not appear for $(c_1, c_2)$ in $\texttt{Quadratic}$, as we expected.

We have emphasized the variation of stochasticity behavior for different galaxy definitions and simulation details. We demonstrate the latter intuitively by comparing the parameters distribution of two sets of simulation samples in Fig.~\ref{fig:add_2} The global features for two sets are similar, but the details in each redshift is different to a large extent. 
\bibliographystyle{apsrev4-2}
\bibliography{citations}

\end{document}